\documentclass[11pt]{article}  

\usepackage{amssymb,amstext,amsmath,amsthm}
\usepackage[dvips]{graphicx}
\usepackage{latexsym}
\usepackage{psfrag}
\usepackage{amsfonts}
\usepackage{bbm} 
\usepackage{color}

\usepackage[all,cmtip]{xy}

\usepackage[totalwidth=450pt, totalheight=590pt, centering]{geometry}

\newcommand{\Ref}[1]{(\ref{#1})}

\newcommand{\be}{\begin{equation}}
\newcommand{\ee}{\end{equation}}
\newcommand{\barray}{\begin{array}}
\newcommand{\earray}{\end{array}}
\newcommand{\bea}{\begin{eqnarray}}
\newcommand{\eea}{\end{eqnarray}}
\newcommand{\bs}{\begin{subequations}}
\newcommand{\es}{\end{subequations}}
\newcommand{\balign}{\begin{align}}
\newcommand{\ealign}{\end{align}}
\newcommand{\equ}{\begin{equation}}
\newcommand{\nequ}{\end{equation}}
\newcommand{\eqa}{\begin{eqnarray}}
\newcommand{\neqa}{\end{eqnarray}}

\newcommand{\bra}[1]{\la {#1}|}
\newcommand{\ket}[1]{|{#1}\ra}
\newcommand{\mean}[1]{\la{#1}\ra}

\def\w{\wedge}
\def\la{\langle}
\def\ra{\rangle}

\newcommand{\R}{\mathbb{R}}

\newcommand{\hh}{{\cal H}}

\newcommand{\cs}{{\cal S}}

\def\to{\rightarrow}

\def\d{\delta}
\def\f{\frac}
\def\tl{\tilde}

\usepackage{verbatim}
\usepackage{bbm}
\def\Id{{\mathbbm 1}}
\def\C{{\mathbbm C}}
\def\R{{\mathbbm R}}

\newcommand{\SU}{\mathrm{SU}}
\newcommand{\SO}{\mathrm{SO}}

\let\eps=\epsilon
\def\th{\theta}
\def\al{\alpha}

\newcommand{\rd}{\mathrm{d}}

\makeatletter
\newcommand*{\simboloG}[1]{%
  \vphantom{\sum}
  \smash{%
    \mathchoice{%
      \raisebox{-.3\height}{\Huge$\m@th\displaystyle#1$}%
      }
      {%
      \raisebox{-.1\height}{\Large$\m@th#1$}%
      }{%
      \raisebox{-.1\height}{\small$\m@th#1$}%
      }{%
      \raisebox{-.1\height}{\LARGE$\m@th#1$}%
      }%
    }}
\newcommand{\BigTimes}{\mathop{\simboloG{\times}}}
\makeatother
\makeatletter

\newcommand*{\simboloB}[1]{%
  \vphantom{\sum}
  \smash{%
    \mathchoice{%
      \raisebox{-.1\height}{\Large$\m@th\displaystyle#1$}%
      }
      {%
      \raisebox{-.1\height}{\Large$\m@th#1$}%
      }{%
      \raisebox{-.1\height}{\small$\m@th#1$}%
      }{%
      \raisebox{-.1\height}{\LARGE$\m@th#1$}%
      }%
    }}

\makeatother

\begin{document}

\title{\bf Polyhedra in loop quantum gravity}
\author{{Eugenio Bianchi$^{a}$, Pietro Don\'a$^{a,b}$ and Simone Speziale$^{a}$}
\smallskip \\ 
{\small $^a$\emph{Centre de Physique Th\'eorique\footnote{Unit\'e Mixte de Recherche (UMR 6207) du CNRS et des Universites Aix-Marseille I, Aix-Marseille II et du Sud Toulon-Var. Laboratoire affili\'e \`a la FRUMAM (FR 2291).}, CNRS-Luminy Case 907, 13288 Marseille Cedex 09, France}} \\
{\small $^b$ \emph{Scuola Normale Superiore, Piazza dei Cavalieri 7, 56126 Pisa, Italy} }
}
\date{\today}

\maketitle

\begin{abstract}
Interwiners are the building blocks of spin-network states. The space of intertwiners is the quantization of a classical symplectic manifold introduced by Kapovich and Millson. Here we show that a theorem by Minkowski allows us to interpret generic configurations in this space as bounded convex polyhedra in $\R^3$: a polyhedron is uniquely described by the areas and normals to its faces. We provide a reconstruction of the geometry of the polyhedron: 
we give formulas for the edge lengths, the volume and the adjacency of its faces. At the quantum level, this correspondence allows us to identify an intertwiner with the state of a quantum polyhedron, thus generalizing the notion of quantum tetrahedron familiar in the loop quantum gravity literature. Moreover, coherent intertwiners result to be peaked on the classical geometry of polyhedra. We discuss the relevance of this result for loop quantum gravity. In particular, coherent spin-network states with nodes of arbitrary valence represent a collection of semiclassical polyhedra. Furthermore, we introduce an operator that measures the volume of a quantum polyhedron and examine its relation with the standard volume operator of loop quantum gravity. We also comment on the semiclassical limit of spinfoams with non-simplicial graphs.
\end{abstract}

\setcounter{tocdepth}{2}
\tableofcontents

\section{Introduction}

Loop quantum gravity (LQG) is a continuous theory, whose Hilbert space is the direct sum of spaces associated to graphs $\Gamma$ embedded in a three-dimensional hypersurface,  ${\cal H}=\oplus_\Gamma {\cal H}_\Gamma$. 
It is often convenient to consider a single graph $\Gamma$, and the associated Hilbert space ${\cal H}_\Gamma$. The truncation captures only a finite number of degrees of freedom of the theory. An important question for us is whether these degrees of freedom can be ``packaged'' as to provide some \emph{approximate} description of smooth 3d geometries \cite{tg,IoCarlo}. We specifically think that it would be useful to have a picture of the classical degrees of freedom captured by ${\cal H}_\Gamma$ in terms of discrete geometries. Such knowledge is for instance relevant for the interpretation of semiclassical states restricted on $\hh_\Gamma$. 

As it turns out, useful insights can be gained looking at the structure of $\hh_\Gamma$. It decomposes in terms of 
SU(2)-invariant spaces ${\cal H}_{F}$ associated to each node of valence $F$. 
For a 4-valent node, it has been known for quite some time that an intertwiner represents the state of a ``quantum tetrahedron'' \cite{Barbieri,BaezBarrett}, namely the quantization of the space of shapes of a flat tetrahedron in $\R^3$ with fixed areas. 
For a generic valence $F$, a natural expectation would be a relation to polyhedra with $F$ faces, as mentioned in \cite{Freidel1} and \cite{tg}. In this paper we clarify the details of this correspondence. 

There are two keys to our result. The first one is the fact that $\hh_F$ is the quantization of a certain classical phase space $\cs_F$, introduced by Kapovich and Millson in \cite{Kapovich}. The second is the fact that there is a unique bounded convex polyhedron with $F$ faces of given areas associated to each point of $\cs_F$.
This is guaranteed by an old theorem by Minkowski \cite{Minkowski}.
The correspondence is up to a measure-zero subset of ``degenerate'' configurations, present also in the 4-valent case.
Accordingly, we have the following relations:

\begin{center}
 polyhedra with $F$ faces $\longleftrightarrow$ classical phase space $\cs_F$ $\longleftrightarrow$ intertwiner space $\hh_F$.
\end{center}

An immediate consequence of these results is a complete characterization of coherent states at a fixed graph: they uniquely define a collection of polyhedra associated to each node of the graph. This provides a simple and compelling picture of the degrees of freedom of $\hh_\Gamma$ in terms of discrete geometries, which are associated with a parametrization of the classical holonomy-flux variables in terms of the twisted geometries introduced in \cite{tg}.

The paper is divided into two parts, concerning respectively the classical geometry of polyhedra, and the notion of quantum polyhedron together with its relevance to loop gravity.
The motivation for the first part comes from the fact that polyhedra have a rich classical geometry. One of the reasons why the notion of quantum tetrahedron has been so fruitful in the developement of loop gravity and spinfoams is the fact that everybody understands the geometry of a classical tetrahedron. To make the extension to higher valence as fruitful, we need first of all to clarify a number of aspects of the geometry of polyhedra.

Minkowski's theorem guarantees that a polyhedron can be reconstructed out of the areas and normals to its faces, just as it happens for the tetrahedron. The new feature here is that there are many possible polyhedra with the same number of faces which differ in their combinatorial structure, i.e. in the adjacency relations of the faces. In the first part of the paper (sections \ref{phase space} and \ref{SecRec}) we focus entirely on the classical geometry of polyhedra, and collect and in some cases adapt various results known in the mathematical literature. We discuss the combinatorial classes of polyhedra, and how the phase space of shapes at given areas can be divided into regions of different classes. We show explicitly how a given configuration of areas and normals can be used to reconstruct the polyhedron geometry, including its edge lengths, volume and combinatorial class. Furthermore, we discuss certain \emph{shape matching conditions} which effectively restrict a collection of polyhedra to (a generalization of) Regge geometries. 

In the second part of the paper (sections \ref{SecLQG} to \ref{SecSF}) we discuss the quantum theory. We first review the construction of the quantization map between the phase space $\cs_F$ and Hilbert space of intertwiners $\hh_F$. This leads to the interpretation of an intertwiner state as the state of a quantum polyhedron, and of coherent intertwiners \cite{LS,CF3} as states describing semiclassical polyhedra. The relevance of polyhedra extend to the whole graph Hilbert space $\hh_\Gamma$, via the twisted geometries variables. The result provides an interpretation of coherent spin-network states in $\hh_\Gamma$ as a collection of semiclassical polyhedra.

Furthermore, we introduce a new operator which measures the volume of a quantum polyhedron. Its definition is based on the knowledge of the classical system behind the intertwiner space $\hh_F$, and has the right semiclassical limit on nodes of any valence. We discuss its relation with the standard volume operator of loop quantum gravity.
Finally, we make some brief remarks on the polyhedral picture, Regge calculus and covariant spin foam models.

\section{The phase space of polyhedra}\label{phase space}

\subsection{Convex polyhedra and Minkowski theorem}

A convex polyhedron is the convex hull of a finite set of points in 3d Euclidean space. It can be represented as the intersection of finitely many half-spaces as
\begin{equation}\label{defPmath}
\mathcal{P} = \left\lbrace x\in \mathbb{R}^3 \, | \, n_i \cdot x \leq h_i, \quad  i=1, \dots, m \right\rbrace,
\end{equation}
where $n_i$ are arbitrary vectors, and $h_i$ are real numbers. 
The abstract description \Ref{defPmath} is non-unique and redundant: the minimal set of half-spaces needed to describe a polyhedron corresponds to taking their number $m$ equal to the number of faces $F$ of the polyhedron. In this paper we are interested in the description of a convex polyhedron with $F$ faces in terms of variables that have an immediate geometric interpretation: the areas of the faces of the polyhedron and the unit normals to the planes that support such faces. 

Let us consider a set of unit vectors $n_i \in \R^3$ and a set of positive real numbers $A_i$ such that they satisfy the \emph{closure} condition
\begin{equation}\label{closure}
C \equiv \sum_{i=1}^F A_i \, n_i = 0.
\end{equation}
In the following, we will refer to this set as ``closed normals''.
A convex polyhedron with $F$ faces having areas $A_i$ and normals $n_i$ can be obtained in the following way. For each vector $n_i$ consider the plane orthogonal it. Then translate this plane to a distance $h_i$ from the origin of $\R^3$. The intersection of the half-spaces bounded by the planes defines the polyhedron, $n_i \cdot x \leq h_i$. We can then adjust the heights so that $h_i = h_i(A)$ so that the faces have areas $A_i$. 
 
Remarkably, a convex polyhedron with such areas and normals always exists. Moreover, it is unique, up to rotations and translations. This result is established by the following theorem due to H. Minkowski \cite{Minkowski,Alexandrov}:

\medskip

\noindent {\bf Theorem} (Minkowski, 1897) 
\begin{enumerate}
\item[(a)] \emph{If $n_1,\ldots,n_F$ are non-coplanar unit vectors and $A_1,\ldots,A_F$ are positive numbers such that the closure condition  
{\rm \Ref{closure}} holds, 
than there exists a convex polyhedron whose faces have outwards normals $n_i$ and areas $A_i$.}
\item[(b)] \emph{If each face of a convex polyhedron is equal in area to the corresponding face with parallel external normal of a second convex polyhedron and conversely, then the two polyhedra are congruent by translation.}
\end{enumerate}
This unicity will play an important role in the following.
Throughout the rest of the paper, we use simply polyhedra to refer to bounded convex polyhedra.

\subsection{Kapovich-Millson phase space as the space of shapes of polyhedra}

Let us consider $F$ vectors in $\R^3$ that have given norms $A_1,\ldots, A_F$ and such that they sum up to zero. The space of such vectors modulo rotations has the structure of a symplectic manifold \cite{Kapovich} and is known as the \emph{Kapovich-Millson phase space}\footnote{In \cite{Kapovich} it is also called the space of shapes of (bended) polygons. To be precise, it is a symplectic manifold up to a finite number of points, corresponding to configurations with one or more consecutive vectors collinear.} $\cs_F$,
\begin{equation}
\textstyle \cs_F=\big\{ n_i \in (S^2){}^F \, | \, \sum_{i=1}^F A_i n_i=0  \big\}/ \SO(3)\;.
\label{Shapes}
\end{equation}
The Poisson structure on this $2(F-3)$-dimensional space is the one that descends via symplectic reduction from the natural SO(3)-invariant Poisson structure on each of the $F$ spheres $S^2$.

\begin{figure}[ht]
\begin{center}
\includegraphics[width=4cm]{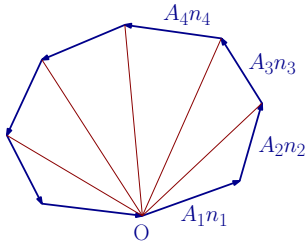} 
\end{center}
\caption{\small \label{FigPol} A polygon with side vectors $A_i n_i$ and the $(F-3)$ independent diagonals. The space of possible polygons in $\R^3$ up to rotations is a $(2F-6)$-dimensional phase space, with action-angle variables the pairs $(\mu_i,\th_i)$ of the diagonal lengths and dihedral angles. For non-coplanar normals, the same data defines also a unique polyhedron thanks to Minkowski's theorem.}
\end{figure}

Action-angle variables for \Ref{Shapes} are $(F-3)$ pairs $(\mu_i, \th_i)$ with canonical Poisson brackets, $\{\mu_i,\th_j\}=\d_{ij}$. Here $\mu_i$ is the length of the  vector $\vec{\mu}_i=A_1n_1+\ldots +A_{i+1}n_{i+1}$ (see Fig.\ref{FigPol}), and its conjugate variable $\th_i$ is the angle between the plane identified by the vectors $\vec{\mu}_{i-1}$, $\vec{\mu}_{i}$ and the plane identified by the vectors $\vec{\mu}_{i}$, $\vec{\mu}_{i+1}$. At fixed areas, the range of each $\mu_i$ is finite. 

Thanks to Minkowski's theorem, a point in $\cs_F$ with non-coplanar normals identifies a unique polyhedron. Accordingly, we refer to \Ref{Shapes} as the \emph{space of shapes of polyhedra at fixed areas}. 
Notice that \Ref{Shapes} contains also configurations with coplanar normals: they can be thought of as ``degenerate'' polyhedra, obtained as limiting cases.
The fact that the polyhedra with faces of given areas form a phase space will be important in section \Ref{quantum polyhedron} where we discuss the Hilbert space of the quantum polyhedron.

\subsection{Classes of polyhedra with $F$ faces}
The phase space $\cs_F$ has a rich structure: as we vary the normals of a polyhedron keeping its areas fixed, not only the geometry, but in general also the combinatorial structure of the polyhedron changes; that is, the number of edges and the adjacency of faces. We refer to the combinatorial structure as the \emph{class} of the polyhedron. In other words, there are two components to the shape of a polyhedron: its class, and its geometry (up to rotations) once the combinatorial structure is fixed. The different classes 
of polyhedra correspond to the different tessellations of a sphere having $F$ faces.
Which class is realized, depends on the specific value of the normals. This is a point we would like to stress: one is not free to choose a class, and then assign the data. It is on the contrary \emph{the choice of data that selects the class}. This is an immediate consequence of Minkowski's theorem.
Accordingly, the phase space $\cs_F$ can be divided into regions corresponding to the different classes of polytopes with $F$ faces. 

\begin{figure}[ht]
\begin{center}
\includegraphics[width=8cm]{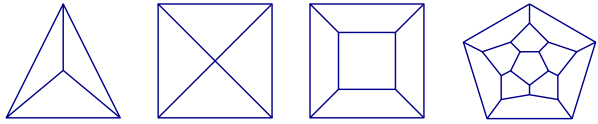}
\end{center}
\caption{\small \label{FigEx} Some examples of Schlegel diagrams. From left to right, a tetrahedron, a pyramid, a cube and a dodecahedron.}
\end{figure}

To visualize the class of a polyhedron it is convenient to use Schlegel diagrams  \cite{Coxeter, Alexandrov}. The Schlegel diagram of a polyhedron is a planar graph obtained choosing a face $f$, and projecting all the other faces on $f$ as viewed from above. See Fig.\ref{FigEx} for examples.

\begin{figure}[ht]
\begin{center}
\includegraphics[width=8cm]{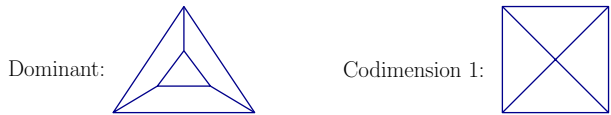}
\end{center}
\caption{\small \label{Fig5} Polyhedra with 5 faces: the two possible classes are the triangular prism (left panel) and the pyramid (right panel). The two classes differ in the polygonal faces and in the number of vertices.
}
\end{figure}

To understand the division of $\cs_F$ into regions of different class, let us first give some examples, and postpone general comments to the end of the Section.
In the most familiar $F=4$ case, there is no partitioning of $\cs_4$: there is a unique tessellation of the sphere, the tetrahedron, and it is well known that there is always a unique tetrahedron associated with four closed normals. The first non-trivial case is $F=5$, where there are two possible classes: a triangular prism, and a pyramid (see Fig. \ref{Fig5}). 
Consider then the phase space $\cs_5$. 
Minkowski's theorem guarantees that the same set $(A_i,n_i)$ cannot be associated to both classes, thus each point in $\cs_5$ corresponds to a unique class. One might at first think that $\cs_5$ can be more or less equally divided among the two classes, but this is not the case. In fact, notice that the pyramid is just a special case of the prism, obtained by collapsing to a point one of the edges connecting two triangular faces. The existence of a pyramid then requires a non-trivial condition, i.e. the presence of a 4-valent vertex.  
A moment of reflection shows that this condition can be imposed via an algebraic equation on the variables. Hence the shapes corresponding to pyramids span a codimension one surface in $\cs_5$.
Generic configurations of areas and normals describe triangular prisms, and the pyramids are measure zero special cases. 
We call \emph{dominant} the class of maximal dimensionality, e.g. the triangular prism here.

\begin{figure}[ht]
\begin{center}
\includegraphics[width=7cm]{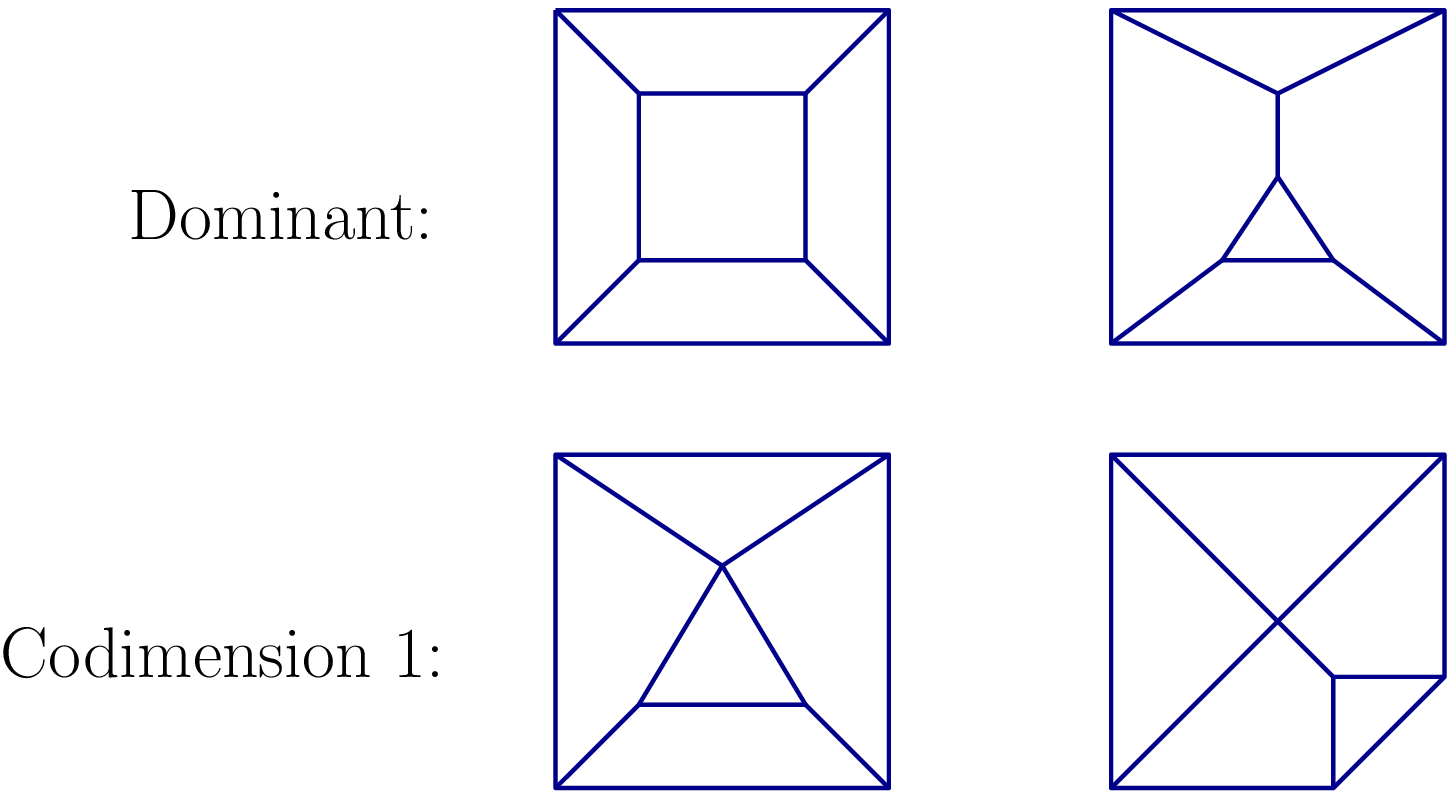} \hspace{1cm} \includegraphics[width=7cm]{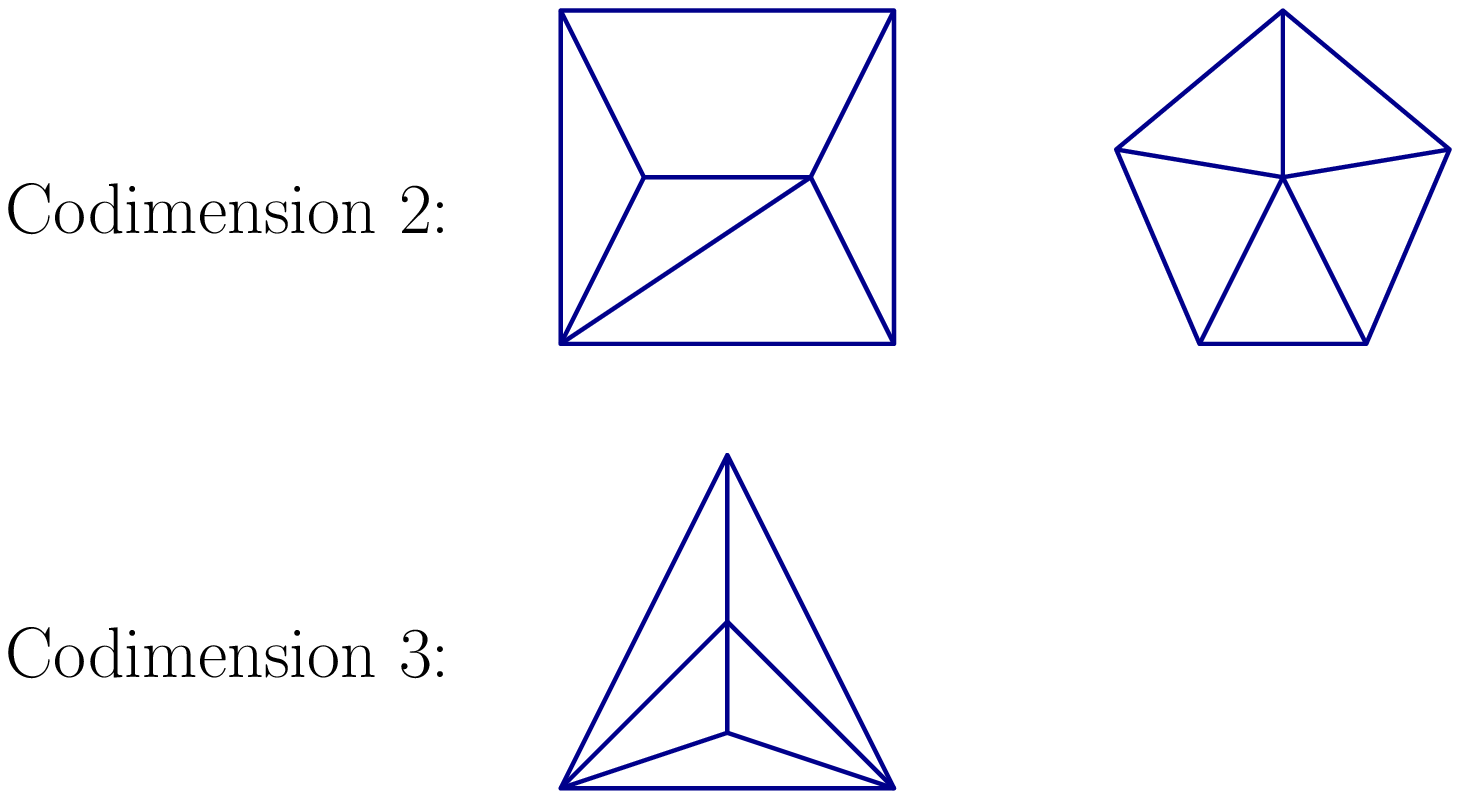}
\end{center}
\caption{\small \label{Fig6} The seven classes of polyhedra with 6 faces, grouped according to the dimensionality of their configurations.}
\end{figure}

Let us move to $F=6$, a case of particular interest since regular graphs in $\R^3$ are six-valent. 
There are \emph{seven} different classes of polyhedra, see Fig.\ref{Fig6}.
The most familiar one is the cuboid (top left of Fig.\ref{Fig6}), with its six quadrilateral faces. 
Remarkably, there is a further dominant class: 
it is a ``pentagonal wedge'', i.e. a polyhedron with two triangles, two quadrilaterals and two pentagons as faces (to visualize it, immagine a triangular prism planed down on a corner, so that a vertex is replaced by a triangle). 
The remaining five classes are subdominant, because non-trivial conditions are required for their existence. 
Subdominant classes have fewer vertices and thus can be seen as special cases with certain edges of zero length.\footnote{Among these, notice the class of codimension 3. It has six triangular faces and three four-valent vertices.
This class is interesting in that it can be seen as two tetrahedra glued along a common triangle. 
Two arbitrary tetrahedra are defined by 12 independent numbers. In order for them to glue consistently and generate this polyhedron, the shape of the shared triangle has to match. 
This shape matching requires three conditions (for instance matching of the edge lengths), thus we obtain a 9-dimensional space of shapes. For fixed external areas, this is precisely the codimension 3 subspace in $\cs_6$.
Hence this class is a special case of two tetrahedra where conditions are imposed for them to glue consistently. 
}

\begin{figure}[ht]
\begin{center}
\includegraphics[width=4cm]{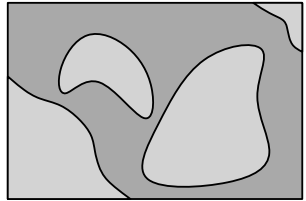}
\end{center}
\caption{\small \label{FigPex} Pictorial representation of the phase space: it can be mapped into regions corresponding to the various dominant classes (two in the example). The subdominant classes separe the dominant ones and span measure-zero subspaces. 
}
\end{figure}

\begin{figure}[h!]
\begin{center}
\includegraphics[scale=.5]{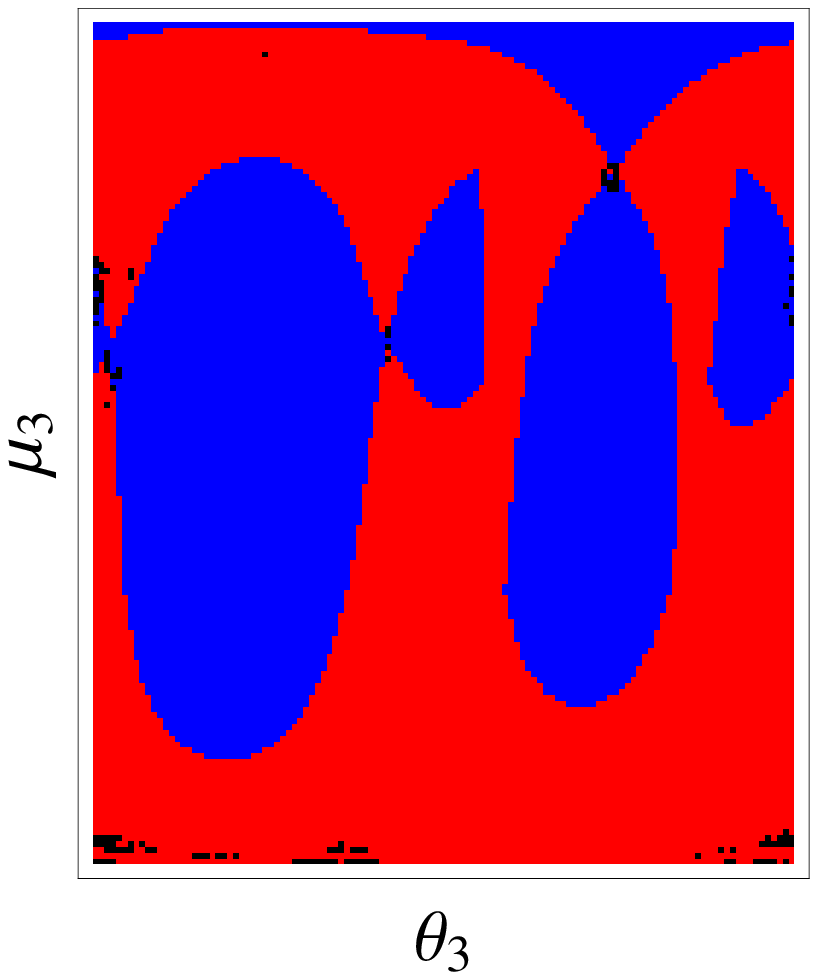} \hspace{2.5cm} \includegraphics[scale=.4]{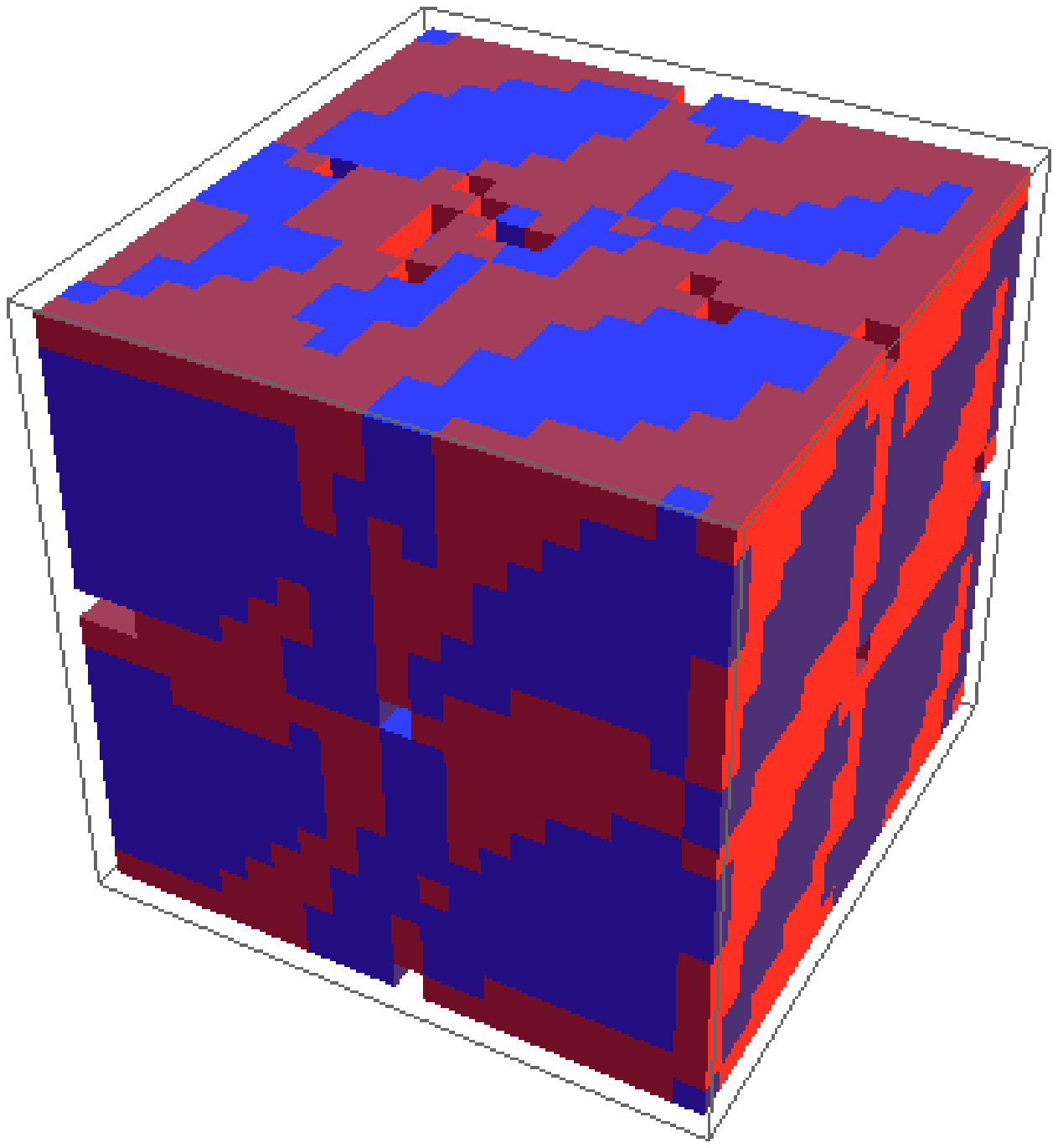} 
\caption{\small{\label{FigMap1} Mappings of subspaces of ${\cal S}_6$ realized using the reconstruction algorithm of Section \ref{SecRec} and Wolfram's Mathematica. We subdivided the phase space into a regular grid, and had Mathematica computing the adjacency matrix of the area-normal configurations lying at the center of the cells. This associates a unique class to each cell of the phase space. The information is colour-coded, cuboids in blue, pentagonal wedges in red.
With this mapping of finite resolution we have measure-zero probability of hitting a subdominant class, thus the latter are absent in the figures. 
The holes are configurations for which our numerical algorithm failed.
Concerning the specific values of the example, the areas are taken to be $(9,10,11,12,13,13)$. In the left panel, we fixed $\mu_1 = 15$, $\th_1 = \frac{7}{10}\pi$, $\mu_2 = 13$, $\th_2 = \frac{13}{10}\pi$, and plotted the remaining  pair $(\mu_3,\th_3)$. In the right panel, we fixed $\mu_i = (15, 13, 17)$ and plotted the three angles $\th_i$. 
}}
\end{center}
\end{figure}

From the above analysis, we expect that the phase space 
$\cs_6$ can then be divided into regions corresponding to the two dominant classes, separed by the subdominant ones. This is qualitatively illustrated in Figure \ref{FigPex}. 
To confirm this picture, we performed some numerical investigations. Using the reconstruction algorithm, which we introduce in the next Section, we can assign a class to each point in $\cs_6$. In Fig.\ref{FigMap1} we give an explicit example of a 2d and a 3d slice of the 6d space $\cs_6$, which shows the subdivision into the two dominant classes.

\medskip

After this brief survey of some specific examples, let us make some general statements.

\begin{itemize}
\item The phase space $\cs_F$ can be divided into regions corresponding to different classes. The dominant classes, generically more than one, cover it densely, whereas the subdominant ones span measure-zero subspaces.
The dominant classes in phase space correspond to polyhedra with all vertices three-valent, that is the dual to the tessellation is a triangulation. This condition maximizes both the number of vertices, $V=3(F-2)$, and edges, $E=2(F-2)$. Subdominant classes are special configurations with some edges of zero lengths and thus fewer vertices.

\item Since all classes correspond to tessellations of the sphere with $F$ faces, they are connected by Pachner moves \cite{Pachner}. 
The reader can easily find a sequence of moves connecting all seven classes of Fig.\ref{Fig6}.
To start, apply a 2-2 move to the upper edge of the inner square of the cuboid to obtain the pentagonal wedge.

\item The lowest-dimensional class corresponds to a maximal number of triangular faces, a condition which minimizes the number of vertices. When \emph{all} the faces are triangular, the polyhedron can be seen as a collection of tetrahedra glued together, and with matching conditions imposed along all shared internal triangles. 

\end{itemize}

\subsubsection{Large $F$ and the hexagonal dominance}

The number of classes grows very fast with $F$ (see for instance \cite{Michon} for a tabulation). 
In the examples above with small $F$, we have been able to characterize the class looking just at how many faces have a certain valence. However as we increase $F$ we find classes with the same valence distribution, but which differ in the way the faces are connected. 
To distinguish the classes one needs to identify the complete combinatorial structure of the polyhedron. This information is captured by the \emph{adjacency matrix}, which codes the connectivity of the faces of the polyhedron. Below in Section \ref{SecAdj} we will show how this matrix can be explicitly built as a function of areas and normals, and give some explicit examples.

\begin{figure}[ht]
\begin{center}
\includegraphics[width=4cm]{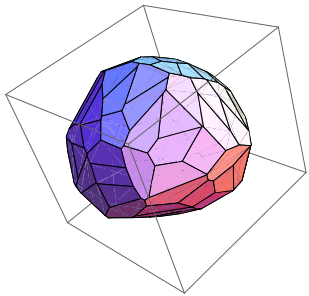}
\end{center}
\caption{\small \label{FigPatata} A polyhedron with $F=100$ drawn with Wolfram's Mathematica, using the reconstruction algorithm. The example has all areas equals and normals uniformly distributed on a sphere. Notice that most faces have valence 6, and that triangles are nowhere to be seen.}
\end{figure}

An interesting question concerns the average valence of a face, defined as $\mean{p}=2E/F$. A simple estimate can be given using the fact that the boundary of any polyhedron is a tessellation of the two-sphere, therefore by the Euler formula $F-E+V=2$. For the dominant classes, which are dual to triangulations, the additional relation $2E= 3V$ holds, hence $E = 3(F-2)$ and we get $\mean{p}=6(1-2/F)$. For large $F$, we expect the polyhedron to be dominated by \emph{hexagonal faces}.
This expectation is immediately confirmed by a simple numerical experiment. 
The specimen in Figure \ref{FigPatata}, for instance, has $F=100$ and $\mean{p}\sim 5.88$.
Notice also from the image that there are no triangular faces, consistently with the fact that they tend to minimize the number of vertices and are thus highly non-generic configurations.

\section{Polyhedra from areas and normals: reconstruction procedure}\label{SecRec}

So far we have discussed how a point in $\cs_F$ specifies a unique polyhedron, and the existence of different combinatorial structures. We now describe how the polyhedron can be explicitly reconstructed from areas and normals. The reconstruction will allow us to evaluate completely its geometry, including the lengths of the edges and the volume, and to identify its class through the adjacency matrix, thus being able to associate a class with each point of $\cs_F$.

The main difficulty in developing a reconstruction algorithm is that, given the areas and the normals, it is not known a priori which faces of the polyhedron are adjacent. The adjacency relations of the faces (and the combinatorial class of the polyhedron) are to be derived together with its geometry. This can be done in two steps. The first step uses an algorithm due to Lasserre \cite{Lasserre} that permits to algebraically compute the lengths $\ell_{ij}(h,n)$ of all the edges of the polyhedron as defined by $h_i$ and $n_i$, as in \Ref{defPmath}.
The second step consists of solving a certain quadratic system to obtain the values of the heights $h_i$ for given areas.

\subsection{Lasserre's reconstruction algorithm}
We now review Lasserre's procedure, and adapt it to the three-dimensional case of interest here.
The basic idea of the reconstruction algorithm is to compute the length of an edge as the length of an interval in coordinates adapted to the edge.
Consider the $i$-th face. From the defining inequalities \Ref{defPmath}, we know that points $x \in \mathbb{R}^{3}$ on this face satisfy
\begin{subequations}\label{defFi}\begin{eqnarray}\label{defFi1}
&& n_i \cdot x = h_i\\ \label{defFi2}
&& n_j \cdot x \leq h_j, \quad i \neq j.
\end{eqnarray}\end{subequations}
We consider the generic case in which $n_i\cdot n_j\neq \pm 1 \ \forall i,j$ (these special configurations can be obtained as limiting cases).
We introduce coordinates $y_i$ adapted to the face, that is
\begin{equation}
n_i \cdot y_i = 0, \qquad y_i = x - (x\cdot n_i) n_i.
\end{equation}
Using \Ref{defFi1} we get $x = h_i n_i + y_i$, which inserted in \Ref{defFi2} gives
\be\label{defr}
y_i \cdot n_j \leq  r_{ij}, \quad i\neq j\;,
\ee
where we have defined
\begin{equation}
r_{ij}\equiv h_j - (n_i \cdot n_j)h_i\;.
\label{eq:rij}
\end{equation}
Hence, the $i$-th face can be characterized either in terms of the $x$ or the $y_i$ coordinates,
\begin{equation}
\label{faceadapted}
\left\lbrace \begin{array}{l}
x \cdot n_i = h_i\\
n_j \cdot x \leq h_j, \quad i \neq j
\end{array} \right.  \quad \longrightarrow \quad
\left\lbrace \begin{array}{l}
y_i \cdot n_i = 0 \\ 
y_i \cdot n_j \leq r_{ij}(h,n), \quad i\neq j
\end{array} \right. 
\end{equation}
Notice that $r_{ij}/\sqrt{1- (n_i\cdot n_j)^2}$ is the distance of the edge $ij$ from the projection of the origin on the $i$-th face.

The next step is to iterate this process and describe an edge in terms of its adapted coordinates. We start from the $i$-th face again, and assume that it is connected to the face $j$, so that the two faces share an edge. Points on the edge $ij$ between the $i$-th and the $j$-th face satisfy 
\begin{eqnarray}
&& y_i \cdot n_i = 0 \\ \label{defEijb}
&& y_i \cdot n_j = r_{ij} \\ \label{defEijc}
&& y_i \cdot n_k \leq r_{ik}, \qquad k\neq i,j.
\end{eqnarray}
 As before, we introduce coordinates $z_{ij}$,  adapted to the edge,
\begin{equation}
n_i \cdot z_{ij} = n_j \cdot z_{ij} = 0, \qquad
z_{ij} = y_i - \left[ n_j - (n_i \cdot n_j) n_i \right] \frac{y_i \cdot n_j}{1 - (n_i \cdot n_j)^2}.
\end{equation}
Using \Ref{defEijb} we get that for a point in the edge
\begin{equation}
\label{edgeCoords}
y_i = \left[ n_j - (n_i \cdot n_j) n_i \right] \frac{h_j - h_i(n_i \cdot n_j)}{1 - (n_i \cdot n_j)^2} + z_{ij}.
\end{equation}
Plugging this in \Ref{defEijc} gives
\begin{equation}
z_{ij} \cdot n_k \leq  b_{ij,k},
\end{equation}
where we have defined 
\begin{equation}
b_{ij,k} \equiv h_k - (n_i \cdot n_k)h_i - \frac{(n_j \cdot n_k) - (n_i \cdot n_j)(n_i \cdot n_k)}{1 - (n_i \cdot n_j)^2}\left[ h_j - h_i(n_i \cdot n_j)\right]\;.
\label{eq:bijk}
\end{equation}

Summarizing as before, going to adapted coordinates the edge is defined by 
\begin{equation} \label{znb}
\left\lbrace \begin{array}{l}
y_i \cdot n_i = 0 \\ y_i \cdot n_j = r_{ij}(h,n) \\ 
y_i \cdot n_k \leq r_{ik}(h,n), \quad k\neq i,j.
\end{array} \right. 
\quad \longrightarrow \quad  \left\lbrace  \begin{array}{l}
z_{ij} \cdot n_i = 0 \\ 
z_{ij} \cdot n_j = 0 \\ 
z_{ij} \cdot n_k \leq b_{ij,k}(h,n), \quad i\neq j\neq k
\end{array} \right. 
\end{equation}

At this point we are ready to evaluate the length of each edge. To that end, we parametrize the $z_{ij}$ coordinate vector in terms of its norm, say $\lambda$, and its direction which is given by the wedge product of the two normals, 
\begin{equation}
z_{ij} = \lambda \frac{n_i \w n_j}{\sqrt{1- (n_i \cdot n_j)^2}}.
\end{equation}
If we define
\begin{equation}
a_{ij,k} \equiv \frac{n_i \w n_j\cdot n_k}{\sqrt{1- (n_i \cdot n_j)^2}},
\end{equation}
we can rewrite the inequalities in \Ref{znb} as
\begin{equation}
\lambda a_{ij,k} \leq b_{ij,k}.
\end{equation}

Finally, the length of the edge is the length of the interval determined by the tightest set of inequalities, i.e. 
\be\label{minmax}
\min_{k | a_{ij,k} >0}\left\{\frac{b_{ij,k}}{a_{ij,k}}\right\}-\max_{k | a_{ij,k} <0}\left\{\frac{b_{ij,k}}{a_{ij,k}}\right\}.
\ee 
Here the minimum is taken over all the $k$'s such that $a_{ij,k}$ is positive, and the maximum over all the $k$'s such that $a_{ij,k}$ is negative. This quantity is symmetric \cite{Lasserre} and satisfies a key property: it can be defined for \emph{any pair of faces} $ij$, not only if their intersection defines an edge in the boundary of the polyhedron, and it is \emph{negative} every time the edge does not belong to the polyhedron \cite{Lasserre}.
Thanks to this property, we can consistently define the edge lengths for any pair of faces $ij$ as
\begin{equation}
\label{edgelength}
\ell_{ij}(h,n) = \max_k \left\lbrace 0, \min_{k | a_{ij,k} >0}\left\{\frac{b_{ij,k}}{a_{ij,k}}\right\}-\max_{k | a_{ij,k} <0}\left\{\frac{b_{ij,k}}{a_{ij,k}}\right\} \right\rbrace .
\end{equation}
The result is a matrix whose entries are the edge lengths (as a functions of the normals and the heights) if the intersection is part of the boundary of the polyhedron, and zero if the intersection is outside the polyhedron.

This formula completes Lasserre's algorithm, and permits one to reconstruct the polyhedron from the set $(h_i,n_i)$. To achieve a description in terms of areas and normals, we need one more step, that is an expression for the heights in terms of the areas.
This can be done using \Ref{edgelength} to compute the areas of the faces. We consider the projection of the origin on the face, and use it to divide the face into triangles. Recall the Lasserre's procedure has provided us with the distance between an edge and the projected origin, see \Ref{faceadapted}. We thus can write
\begin{equation}
\label{areas}
A_i = \frac{1}{2} \sum_{\substack{j=1\\j\neq i}}^{F}
\frac{r_{ij}}{\sqrt{1- (n_i\cdot n_j)^2}} \, \ell_{ij}.
\end{equation}
Notice that both $r_{ij}(h,n)$ from \eqref{eq:rij} and $\ell_{ij}(h,n)$ from \eqref{edgelength} are linear in the heights. Hence, the area is a quadratic function,
\begin{equation}\label{Adih}
A_i(h,n) = \sum_{j,k=1}^F M_i^{jk}(n_1, \ldots, n_F) h_j h_k,
\end{equation}
where $M_i$ is a matrix depending only on the normals. 
This homogeneous quadratic system can be solved for $h_i(A,n)$. 
The existence of a solution with $h_i>0$ $\forall i$ is guaranteed by Minkowski's theorem. 
However, the solution is not unique: in fact, we have the freedom of moving the origin around inside the polyhedron, thus changing the value of the heights without changing the shape of the polyhedron. A method which we found convenient to use is to determine a solution minimizing the function
\be
f(h_i) \equiv \sum_i \left( A_i(h,n) - A_i \right)^2
\ee
at areas and normals fixed, with $A_i(h,n)$ given by \Ref{Adih}. This is the method used in the numerical investigations of Figs. \ref{FigMap1} and \ref{FigPatata}.\footnote{Concerning Fig. \ref{FigMap1}, we can also give now more details on the holes: these are
configurations for which the numerical algorithm to solve \Ref{Adih} failed. This limitation can be easily improved with a better inversion algorithm, or by choosing a configuration slighly off the center of the cell.
}

Finally, from the inverse we derive the lengths as functions of areas and normals, which with a slight abuse of notation we still denote in the same way,
\be
\ell_{ij}(A,n) = \ell_{ij}(h(A,n),n).
\ee
These expressions are well-defined and can be computed explicitly.

\subsection{Volume of a polyhedron in terms of areas and normals}\label{SecV}

Let us call $\mathcal{P}(A_i,n_i)$ the convex subset of $\mathbb{R}^3$ corresponding to the polyhedron. Its volume is simply the integral on this region of the Euclidean volume density:
\begin{equation}
V(A_i,n_i)=\int_{\mathcal{P}(A_i,n_i)}\hspace{-1.5em}d^3 {x} .
\label{eq:Vint}
\end{equation}
An interesting question is how to compute efficiently the volume integral \Ref{eq:Vint}. The simplest way is to use the algorithm described in the previous section: we chop the region $\mathcal{P}(A_i,n_i)$ into pyramids with a common vertex in its interior and bases given by the faces of the polyhedron. In this way the volume is just the sum of the volumes of the pyramids, i.e.
\begin{equation}
\label{volumepiramidi}
V(A_i,n_i)= \frac{1}{3} \sum_{i=1}^F h_i\, A_i\;.
\end{equation}
Here $h_i=h_i(A,n)$ are the heights of the pyramids expressed in terms of the areas and normals via Lasserre's algorithm. 

The volume can be used to define a volume function on the phase space $\cs_F$. To that end, notice that \Ref{volumepiramidi} is not defined for configurations with coplanar normals, which on the other hand do enter $\cs_F$. However, it can be straightforwardly extended to a function on the whole $\cs_F$ by defining it to be zero for coplanar configurations. Furthermore, the resulting phase space function is continuous.\footnote{In order to to see this, one shows that the limit of coplanar normals exists and the volume tends to zero in this limit. From property (C3) -- see below, a general $F$-valent coplanar configuration can be obtained from a $F+1$ configuration in the limit of zero base's area. }
Since the volume is manifestly invariant under rotations, it can also be written as a function of the reduced phase space variables  only, that is, $V(A_i,\mu_k,\th_k)$. To do so explicitly, one uses the relation $n_i=n_i(\mu_k,\th_k)$, which is straightforward to derive once a reference frame is chosen.

The volume of the polyhedron as a function of areas and normals has a number of interesting properties:
\begin{itemize}
	\item[C1.] \emph{Non-negative phase-space function.} The volume is by construction non-negative, and at given areas, it vanishes only when the normals $n_i$ lie in a plane. 
This in particular implies that the volume vanishes for $F=2$ and $3$.

	\item[C2.] \emph{Boundedness.} For fixed areas $A_i$, the volume is a bounded function of the normals. We call $V_{\text{max}}(A_i)$ the volume of the polyhedron with maximum volume,\footnote{Notice that there can be more than one polyhedron that attains maximum volume. For instance, in the case $F=4$, there are two parity-related tetrahedra with maximal volume.}
\begin{equation}
V_{\text{max}}(A_i)\equiv \sup_{n_i} \{V(A_i,n_i)\}\;.
\label{eq:Vmax}
\end{equation}
In particular, $V_{\text{max}}(A_i)$ is smaller that the volume of the sphere that has the same surface area as the polyhedron. Therefore we have the bound
\begin{equation}
0\leq V(A_i,n_i)< \frac{\big(\sum_i A_i\big)^\frac{3}{2}}{3\sqrt{4\pi}}\;.
\label{eq:Vbound}
\end{equation} 
	\item[C3.] \emph{Face-consistency}. If we set to zero one of the areas such that the result is still a non-degenerate polyhedron, the function \Ref{volumepiramidi} automatically measures the volume of the reduced polyhedron with $F-1$ faces.
\end{itemize}

\medskip

In conclusion, a point in $\cs_F$ determines uniquely the whole geometry of a polyhedron and in particular its edge-lengths $\ell_{ij}$ \Ref{edgelength} and its volume \Ref{volumepiramidi}.\footnote{It is worth adding that the problem of computing the volume of a given polyhedron is a complex and well studied topic in computational mathematics \cite{VolumeComplexity, VolumeReview}, hence better procedures than the one used here could in principle be found. However, the usual starting point for common algorithms is the knowledge of the coordinates of vertices, or the system of inequalities \Ref{defPmath}. Therefore the methods need to be adapted to obtain formulas in terms of areas and normals. 
The main difficulty is clearly that the adjacency relations of the faces are to be derived together with the geometry. We found Lasserre's algorithm to be the most compatible with these necessities, thanks to the fact that the lengths are reconstructed algebraically. Numerical algorithms for the volume and shape reconstruction from areas and normals are developed in the study of extended Gaussian images in informatics \cite{Little}, however there are no analytical results.} 
Now we show how these data can be used to identify the class of the polyhedron.

\subsection{Adjacency matrix and the class of the polyhedron}\label{SecAdj}

The adjacency matrix $A$ of the polyhedron is defined as 
\begin{equation}\label{defA}
\quad A_{ij}=\left\lbrace\begin{array}{cl}
1 & \text{if the faces $i$ and $j$ are adjacent } \\ 
0 & \text{otherwise}
\end{array}\right. \quad i,j = 1,\ldots,F
\end{equation}
Notice that $A_{ij}$ coincides with the matrix $\ell_{ij}$ in \Ref{edgelength} with all the non-zero entries normalized to $1$: the recontruction algorithm gives us the adjacency matrix for free.

The symmetric matrix $A_{ij}$ contains information on the connectivity of the faces as well as on the valence of each face, thus the class of the polyhedron can be  identified uniquely from it. The valence $p_i$ of the face $i$ can be extracted taking the sum of the columns for each row,
\begin{equation}
p_i = \sum_{j=1}^F A_{ij}.
\end{equation}
For example, for the two classes with $F=5$ of Fig.\ref{Fig5} we have

$$
\begin{picture}(0,0)(0,0)\put(-65,-20){\includegraphics[scale=.5]{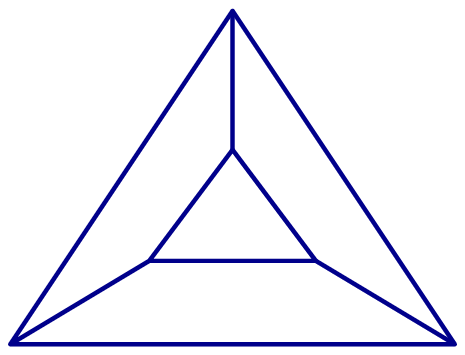}}\end{picture}
\ \longrightarrow \ \ A = \left(\begin{array}{ccccc}
0 & 0 & 1 & 1 & 1 \\
0 & 0 & 1 & 1 & 1 \\
1 & 1 & 0 & 1 & 1 \\
1 & 1 & 1 & 0 & 1 \\
1 & 1 & 1 & 1 & 0 \\
\end{array}\right) \ \longrightarrow \ \ 
p = \left( 3, 3, 4, 4, 4 \right)
$$

$$
\begin{picture}(0,0)(0,0)\put(-55,-20){\includegraphics[scale=.5]{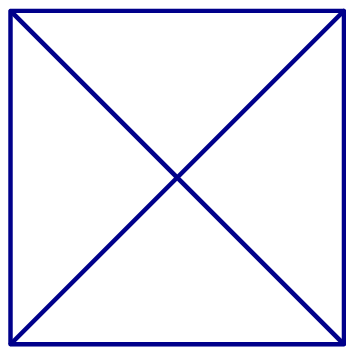}}\end{picture}
\ \longrightarrow \ \ A = \left(\begin{array}{ccccc}
0 & 1 & 1 & 1 & 1 \\
1 & 0 & 1 & 0 & 1 \\
1 & 1 & 0 & 1 & 0 \\
1 & 0 & 1 & 0 & 1 \\
1 & 1 & 0 & 1 & 0 \\
\end{array}\right) \ \longrightarrow \ \ 
p = \left( 4, 3, 3, 3, 3 \right)
$$

From graph theory \cite{Godsil}, we known that \Ref{defA} has a number of interesting properties that can be related to the geometrical parameters of the polyhedron. For instance, the number of walks from the face $i$ to the face $j$ of length $r$ is given by the matrix elements of the $r$-th power $(A^r)_{ij}$. From this property we deduce that the number $E$ of edges of the polyhedron is
\begin{equation}
E = \f12 \, \mathrm{Tr} A^2 = \f12 \sum_i p_i.
\end{equation}
This expression generalizes the value $E=3(F-2)$ valid for the dominant classes.

Higher traces are related to the number of loops of a given lengh. For instance, the number of closed loops of length 3 is given by 
$(1/6) \,\mathrm{Tr} A^3.$

Through the adjacency matrix, obtained via the reconstruction procedure, areas and normals identify a unique class, and thus permits the division of $\cs_F$.

\subsection{Shape-matching conditions}\label{SecSM}

Knowing the complete geometry of the polyhedra allows us also to address the following situation.
Suppose that we are given two polyhedra in terms of their areas and normals, and that we want to glue them by a common face. Even if we choose the area of the common face to be the same, there is no guarantee that the shape of the face will match: The two sets of data will in general induce different shapes of the face. 
That is, the face has the same area but it can be two different polygons altogether.
In order to glue the polyhedra nicely, one needs shape matching conditions guaranteeing that the shared face has the same geometry in both polyhedra.

\begin{figure}[h!]\begin{center}
\includegraphics[width=2cm]{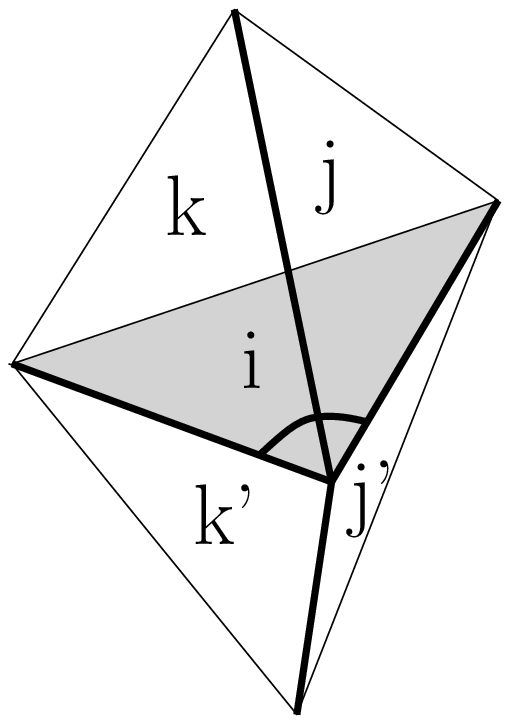} \hspace{3cm} \includegraphics[width=3cm]{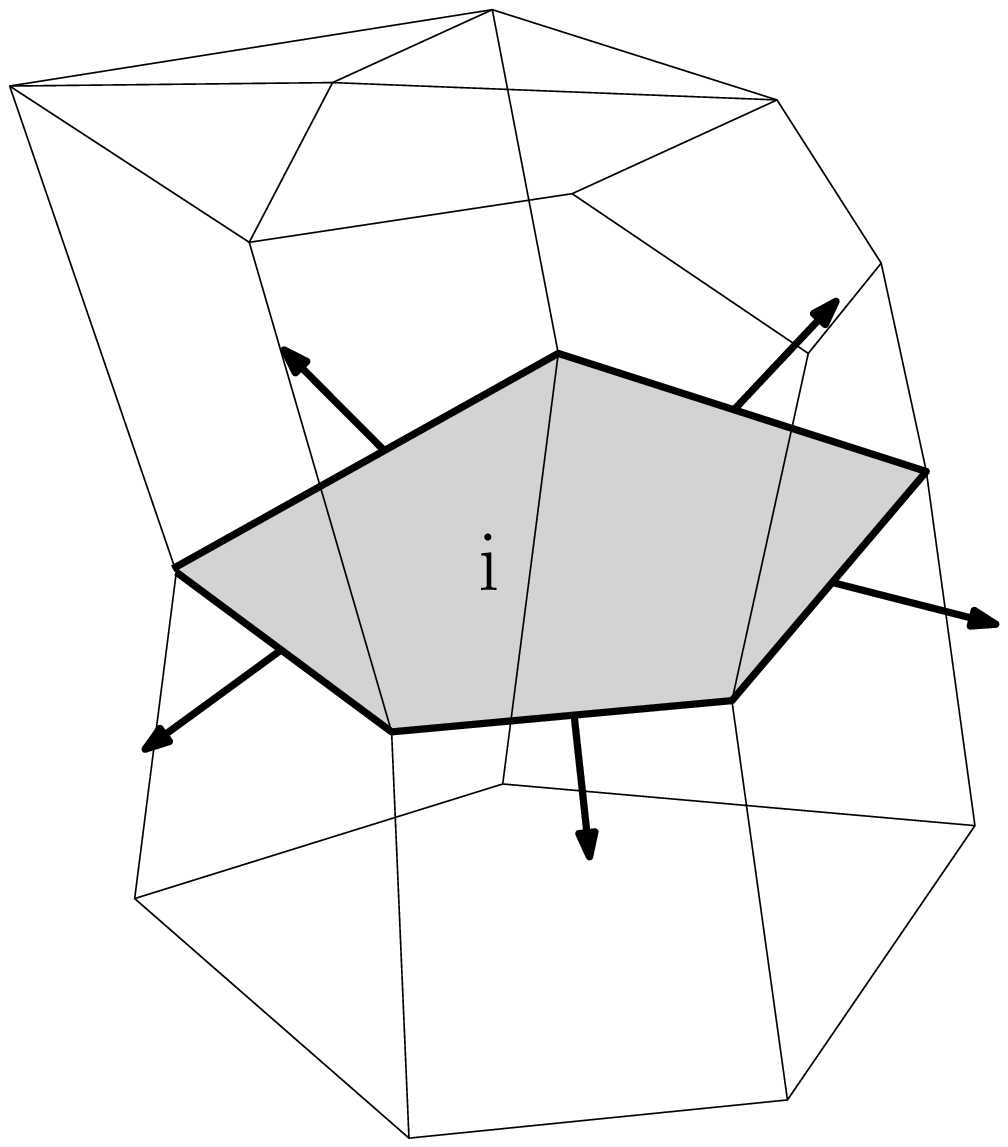}
\end{center}
\caption{\label{Figaa}\small{ The geometric meaning of equation \Ref{aa}: the 2d angle $\al_{ij,kl}$ belonging to the shaded
triangle can be expressed in terms of 3d angles associated the thick edges of the tetrahedron $k$, or
equivalenty of the tetrahedron $l$.}}
\end{figure}

If both polyhedra are tetrahedra, the problem has been solved in \cite{Dittrich}.
One uses the fact that the shape of the common triangle matches if two lengths, or two internal angles, are the same. 
The internal angles $\alpha$ can be expressed in terms of the 3d dihedral angles of the tetrahedron as follows,
\eqa\label{a}
\cos\alpha^i_{jk}=\frac{\cos\phi_{ij}+\cos\phi_{ik} \cos\phi_{jk}  }{\sin\phi_{ik} \sin\phi_{jk}}.
\neqa
Here the faces $i$, $j$ and $k$ all share a vertex, and $\al^i_{jk}$ is the angle between the edge $ij$ and the edge $ik$ inside the triangle $i$. 
Consider now the adjacent tetrahedron. Its geometry induces for the same angle the value
\eqa\label{a1}
\cos\alpha^i_{j'k'}=\frac{\cos\phi'_{ij'}+\cos\phi'_{ik'} \cos\phi'_{j'k'}  }{\sin\phi'_{ik'} \sin\phi'_{jk'}}.
\neqa
Hence, for the shape to match it is sufficient to require
\be\label{aa}
{\cal C}_{kl,ij}(\phi) \equiv \cos\al^i_{jk} - \cos\al^i_{j'k'} = 0
\ee
for two of the three angles of the triangle. These \emph{shape matching conditions} are conditions on the normals of the two tetrahedra.
See left panel of Figure \ref{Figaa} for an illustration of these relations.

The simplicity of the conditions \Ref{aa} is a consequence of the fact that two triangles with the same area are congruent if two angles match. For the general case, the face to glue is now a polygon and the number of conditions greater. One needs to make sure that the valence $p$ of the polygon is the same. Then, the number of independent parameters of a polygon on the plane is $2p-3$, hence giving the edge lengths is not enough, and $p-2$ additional conditions are needed. 
A convenient procedure is the following. Identify the faces of the two polyhedra that, having the same area, we want to match. From the reconstruction algorithm, we know the edge lengths $\ell_{ij}$ of the face viewed from one polyhedron. Then, for all $j$ such that $\ell_{ij}\neq 0$, we consider the face normals $n_j$ projected on the plane of the $i$-th face,
\be
\tl n_j =\f{n_j - (n_i\cdot n_j) n_i}{|n_j - (n_i\cdot n_j) n_i|} = 
\f{n_j - \cos\phi_{ij} n_i}{\sin\phi_{ij}}.
\ee
The set $(\ell_{j}, \tl n_j)$ defines a unique polygon in the plane identified by $n_i$, thanks to a two-dimensional version of Minkowski's theorem.
Then, we do the same with the second polyhedron, obtaining a second set $(\ell_{j}', \tl n_j')$ living in the plane identified by $n_i'$. Finally, the shape matching conditions consist of imposing the equivalence of these two flat polygons up to rotations in three-dimensional space. Notice that the shape matching are now conditions on both the normals and the areas of the two polyhedra.

\section{Relation to loop quantum gravity}\label{SecLQG}
Thus far we have been discussing classical properties of polyhedra. In the rest of the paper, we discuss the relevance of polyhedra for loop quantum gravity. The relation comes from the following two  key results:
\begin{itemize}
	\item[(\emph{i})] Intertwiners are the building blocks of spin-network states, an orthonormal basis of the Hilbert space of loop quantum gravity \cite{RS1,Baez:1994hx}
	\item[(\emph{ii})] Intertwiners are the quantization of the phase space of Kapovich and Millson \cite{Charles,CF3,FKL} (see also \cite{Roberts,KapoQ}),  i.e.  of the space of shapes of polyhedra with fixed areas discussed in the previous sections. 
\end{itemize} 
Therefore an intertwiner can be understood as the state of a \emph{quantum polyhedron}, and spin-network states as a collection of quantum polyhedra associated with each vertex.

In this section we review how $(ii)$ and the notion of quantum polyhedron are established, observe that coherent intertwiners are peaked on the geometry of a classical polyhedron and discuss the relevance of this fact for the relation between semiclassical states of loop quantum gravity and twisted geometries.

\subsection{The quantum polyhedron}\label{quantum polyhedron}

Let us consider the space of vectors in 3d Euclidean space with norm $j$. This is a phase space, the Poisson structure being the rotationally invariant one proper of the 2-sphere $S^2_j$ of radius $j$. As is well known, its quantization\footnote{Notice that, as usual, the quantum theory requirers the quantization of some classical quantities. In this case the norm of the vector has to be a half-integer $j$, the \emph{spin}.} is the representation space $V^{(j)}$ of SU(2). We are interested in the phase space $\cs_F$, that is the space of $F$ vectors that sum to zero, up to rotations. The Poisson structure on $\cs_F$ is obtained via the symplectic reduction of the Poisson structure on the product of $F$ spheres of given radius. Thanks to Guillemin-Sternberg's theorem that quantization commutes with reduction,\footnote{For the general theory see \cite{GS}, for details on the application to the current system see 
\cite{BaezBarrett} and in particular~\cite{CF3}.}
we can quantize first the unconstrained phase space $\times_iS^2_{j_i}$, and then reducing it at the quantum level extracting the subspace of $\otimes_iV^{(j_i)}$ that is invariant under rotations. 
This gives precisely the intertwiner space $\hh_F = {\rm Inv} \big[\otimes_{i=1}^F V^{(j_i)}\big]$. The situation is summarized by the commutativity of the following diagram,

\begin{center}
\begin{tabular}{ccccc}
& $\times_i S^2_{j_i}$ & $\longrightarrow$ & $\otimes_i V^{j_i}$ & \\
{\footnotesize{Symplectic reduction}} & $\downarrow$ & & $\downarrow$ & {\footnotesize{Quantum reduction}} \\
& ${\cs}_F$ & $\longrightarrow$ & ${\cal H}_F $ & 
\end{tabular}
\end{center}

The correspondence between classical quantities and their quantization is the following: up to a dimensionful constant, the generators $\vec{J}_i$ of SU(2) acting on each representation space $V^{(j_i)}$ are understood as the quantization of the vectors $A_i n_i$. In LQG the dimensionful constant is chosen to be the Immirzi parameter $\gamma$ times Planck's area $8\pi L_P^2$,
\begin{equation}
A_i n_i\qquad \longrightarrow\qquad \hat{E}_i=8\pi \gamma L_P^2\;\vec{J}_i.
\label{eq:corresp}
\end{equation}
The closure condition (\ref{closure}) on the normals of the polyhedron is promoted to an operator equation,
\begin{equation}
\sum_{i=1}^F\vec{J}_i\,=0.
\label{closureQ}
\end{equation}
This condition defines the space of intertwiners, and corresponds to the Gauss constraint of classical General Relativity in Ashtekar-Barbero variables. 

One can then proceed to associate operators to geometric observables through the quantization map \Ref{eq:corresp}. The area of a face of the quantum polyhedron is 
\begin{equation}
\hat{A}_i=\sqrt{\hat{E}_i\cdot\hat{E}_i}=8\pi \gamma L_P^2 \sqrt{j_i(j_i+1)}
\label{eq:area}
\end{equation}
and produces an equispaced quantization of the area $A_i\sim j_i$ for large spins, i.e. up to quantum corrections.
Notice that an ordering can be chosen so that the area is exactly $\hat{A}_i=8\pi \gamma L_P^2 j_i$. 
This ordering will be considered below to simplify the construction of the volume operator.

The scalar product between the generators of $SU(2)$ associated to two faces of the polyhedron measures the angle $\theta_{ij}$ between them \cite{Major},
\begin{equation}
\hat{\theta}_{ij}=\arccos\frac{\vec{J_i}\cdot\vec{J_j}}{\sqrt{j_i(j_i+1)\;j_j(j_j+1)}}.
\label{eq:angle}
\end{equation}
Notice that the angle operators do not commute among themselves, therefore it is not possible to find a state for a quantum polyhedron that has a definite value of all the angles between its faces. Moreover, the adjacency relations of the faces is not prescribed a priori,
thus $\hat{\theta}_{ij}$ might not even be a true dihedral angle of the polygon. 
Therefore an eigenstate of a maximal commuting set of angles is far from the state of a classical polyhedron: it is an infinite superposition of polyhedra of different shapes, including different combinatorial classes. Semiclassical states for a quantum polyhedron are discussed in the next section.

\subsection{Coherent intertwiners and semiclassical polyhedra}
Coherent intertwiners for $\hh_F$ were introduced in \cite{LS} and furtherly developed in \cite{CF3,FKL} (for previous related work, see \cite{RovelliSemi}). These Livine-Speziale (LS) coherent intertwiners are defined as the SU(2)-invariant projection of a tensor product of states $\ket{j_i,n_i}\in V^{(j_i)}$,
\be
|\!\ket{j_i,n_i} \equiv \int \rd g \ D^{(j_1)}(g) \ket{j_1,n_1}\cdots D^{(j_F)}(g)\ket{j_F,n_F}.
\label{CI}
\ee
The states $\ket{j,n}$ are SU(2) coherent states peaked on the direction $n$ of the spin \cite{Perelomov, Klauder},
\begin{equation}
\bra{j,n}\vec{J}\ket{j,n}=j n.
\label{eq:jn}
\end{equation}
In (\ref{CI}), the unit-vectors $n_i$ can be assumed to close, $\sum_i j_i n_i=0$. The reduced states are still an overcomplete basis of $\hh_F$, as a consequence of the Guillemin-Sternberg theorem \cite{CF3,Barrett}.

Coherent intertwiners are semiclassical states for a quantum polyhedron: the areas are sharp, and the expectation value of the non-commuting angle operators $\hat{\theta}_{ij}$ reproduces the classical angles between faces of the polyhedron in the large spin limit,
\begin{equation}
\frac{\bra{j_i,n_i}\!|\cos \hat{\theta}_{ij}|\!\ket{j_i,n_i}}{\bra{j_i,n_i}\!\ket{j_i,n_i}} \approx n_i\cdot n_j.
\label{eq:<theta>}
\end{equation}
Moreover, the dispersions are small compared to the expectation values.

A useful fact is that coherent intertwiners can be labeled directly by a point in the phase space $\cs_F$ of Kapovich and Millson, and therefore by a unique polyhedron. This provides a resolution of the identity in intertwiner space as an integral on $\cs_F$.
To realize this reduction, it is convenient to parametrize $\cs_F$ via $F-3$ complex numbers $Z_k$, instead of $(\mu_k,\th_k)$.  
Let us choose an orientation in $\R^3$ and consider the stereographic projection $z_i$ of the unit-vectors $n_i$ into the complex plane.\footnote{The relation between the unit-vector $n=(n_x,n_y,n_z)$ and the stereographic projection is 
\be
z = - \f{n_x-i n_y}{1-n_z} = -\tan\f\th2 \, e^{-i\phi},\nonumber
\ee
where $\th$ and $\phi$ are the zenith and azimuth angles of $S^2$, and we have chosen to project from the south pole.} The $F-3$ complex variables $Z_k$ are the cross-ratios \cite{CF3}
\begin{equation}
Z_k=\frac{(z_{k+3}-z_1)(z_2-z_3)}{(z_{k+3}-z_3)(z_2-z_1)}, \quad k=1,\ldots, F-3.
\label{eq:crossratio}
\end{equation} 
Given an orientation in $\R^3$, a set of normals $n_i$ that satisfy the closure condition \Ref{closure} can be obtained as a function of the cross-ratios, 
\begin{equation}
n_i=n_i(Z_k)\;.
\label{eq:nZ}
\end{equation}
Coherent intertwiners can then be obtained via geometric quantization \cite{CF3}: they are labeled by the variables $Z_k$, that is $\ket{j_i,Z_k}$, and are equal to the states $|\!\ket{j_i,n_i}|=\!\ket{j_i,n_i(Z_k)}$ up to a normalization and phase.\footnote{The states $\ket{j_i,Z_k}$ also define an holomorphic representation of the quantum algebra of functions $\psi(Z_k)\equiv \bra{j_i,\bar Z_k}\psi\ra$, see \cite{FKL}. We will not use this representation in this paper.
}
The resolution of the identity is given by an integral over the variables $Z_k$,
\begin{equation}
\Id_{\hh_F}=\int_{\C^{F-3}} \rd\mu(Z_k)\; \ket{j_i,Z_k}\bra{j_i,Z_k}\;,
\label{eq:res id}
\end{equation}
where the integration measure $\rd\mu(Z_k)=K_{j_i}(Z_k, \bar Z_k)  \prod_k \rd^2 Z_k$ depends parametrically on the spins $j_i$ and is given explicitly in \cite{CF3}. The relevance of this formula for the following discussion is that it provides a resolution of the identity in intertwiner space as a sum over semiclassical states, each one representing a classical polyhedron: the intertwiner space can be fully described in terms of polyhedra.\footnote{Recently \cite{Freidel1,Freidel2,Borja} attention has been given to a second space for which polyhedra are relevant. This is a sum of intertwiner spaces such that the total spin is fixed,
\be\label{hhJ}
\hh_J = \underset{\underset{\sum_ij_i = J} {j_1 .. j_F}}\oplus {\rm Inv} \big[\otimes_{i=1}^F V^{(j_i)}\big].
\ee
The interest in this space is that it is a representation of the unitary group U(F). Vectors in this space represent quantum polyhedra with fixed number of faces and fixed total area, but fuzzy individual areas as well as shapes as before. 
Coherent states for \Ref{hhJ} can be built using U(F) coherent states \cite{Freidel2}. These are also peaked on classical polyhedra like the LS states \Ref{CI}, thus the results in this paper are relevant for them as well.}

\subsection{Coherent states on a fixed graph and twisted geometries}\label{twisted}

The states $\ket{j_i,Z_k}$ provide coherent states for the space of intertwiners only,
and should not be confused with coherent spin-network states for loop quantum gravity. Nevertheless, classical polyhedra and coherent intertwiners are relevant to the full theory, as we now discuss.

To relate polyhedra to loop quantum gravity, consider a truncation of the theory to a single graph $\Gamma$, with $L$ links and $N$ nodes. The associated gauge-invariant Hilbert space ${\cal H}_\Gamma = L_2[\SU(2)^L/\SU(2)^N]$ decomposes in terms of intertwiner spaces $\hh_{F(n)}\equiv {\rm Inv}[\otimes_{l \in n} V^{(j_l)}]$ as
\be\label{Hs1}
{\cal H}_\Gamma = \oplus_{j_l} \left(\otimes_n {\cal H}_{F(n)} \right).
\ee 
This Hilbert space is the quantization of a classical space\footnote{Again, this is a symplectic manifold up to singular points \cite{Bahr}.}
$S_\Gamma = T^*\SU(2){}^L /\!/ \SU(2)^N$, which corresponds to (gauge-invariant) holonomies and fluxes 
associated with links and dual faces of the graph. The double quotient $/\!/$ means symplectic reduction.
The key result is that this space admits a decomposition analogous to \Ref{Hs1}. In fact, it can be parametrized as 
the following Cartesian product \cite{tg},
\be\label{S}
S_\Gamma = \BigTimes_l \, T^*S^1 \BigTimes_n \cs_{F(n)} ,
\ee
where $T^*S^1$ is the cotangent bundle to a circle, $F$ the valence of the node $n$, and $\cs_F$ is the phase space of Kapovich and Millson.

The parametrization is achieved through an isomorphism between holonomy-fluxes and a set of variables dubbed ``twisted geometries''. These are the assignment of an area $A_l$ and an angle $\xi_l$ to each link, and of $F$ normals $n_i$, satisfying the closure condition \Ref{closure}, to each node. 
See \cite{tg,IoCarlo} for details and discussions. In this parametrization, a point of $S_\Gamma$ describes a collection of polyhedra associated to each node. The two polyhedra belonging to nodes connected by a link $l$ share a face. The area of this face is uniquely assigned to both polyhedra $A_l$ (notice that this fact alone does not imply that the shape of the face matches -- more on this below).
The extra angles $\xi_l$ carry information on the extrisic geometry between the polyhedra. 

The isomorphism \Ref{S} and the unique correspondence between closed normals and polyhedra means that \emph{each classical
holonomy-flux configuration on a fixed graph can be visualized as a collection of polyhedra, together with a notion of parallel transport between them.}
Just as the intertwiners are the building blocks of the quantum geometry of spin networks, polyhedra are the building blocks of the classical phase space \Ref{S} in the twisted geometries parametrization. 
\\

What is the relevance of this geometric construction to the quantum theory? 
Coherent states for loop quantum gravity have been introduced and extensively studied by Thiemann and collaborators \cite{Thiemann,Hall,Bahr}. Although the states for the full theory have components on each graph, one needs to cut off the number of graphs to make them normalizable. In practice, it is often convenient to truncate the theory to a single graph. This truncation provides a useful computational tool, to be compared to a perturbative expansion, and has found many applications, from the study of propagators \cite{Bianchi:2009ri} to cosmology \cite{BRVcosmo}. In many of these applications, control of the semiclassical limit requires a notion of semiclassical states in the truncated space ${\cal H}_\Gamma$.
The truncation can only capture a finite number of degrees of freedom, thus coherent states in ${\cal H}_\Gamma$ are not peaked on a smooth classical geometry. Twisted geometries offer a way to see them as peaked on a discrete geometry, to be viewed as an approximation of a smooth geometry on a cellular decomposition dual to the graph $\Gamma$. The above results provide a compelling picture of these twisted geometries in terms of polyhedra, and thus of coherent states as a collection of semiclassical polyhedra.

There is one subtlety with this geometric picture that should be kept in mind, which justifies the name ``twisted'' geometries: they define a metric which is locally flat, but \emph{discontinuous}. To understand this point, consider the link shared by two nodes. Its dual face has area proportional to $A_l$. However, the shape of the face is determined \emph{independently} by the data around each node (i.e. the normals and the other areas), thus generic configurations will give two different shapes. In other words, the reconstruction of two polyhedra from holonomies and fluxes does not guarantee that the shapes of shared faces match. Hence, the metric of twisted geometries is discontinuous across the face \cite{tg,IoCarlo}.\footnote{Aspects of this discontinuity have been discussed also in \cite{Eugenio,BiancaJimmy}}
See left panel of Figure \ref{Figaa}.

One can also consider a special set of configurations for which the shapes match, see right panel of Figure \ref{Figaa}. This is a subset of the phase space $S_\Gamma$ where the shape matching conditions, discussed earlier in Section \ref{SecSM}, hold.
This subset corresponds to piecewise flat \emph{and continuous} metrics. For the special case in which all the polyhedra are tetrahedra, this is the set-up of Regge calculus, and those holonomies and fluxes indeed describe a 3d Regge geometry: twisted geometries with matching conditions amount to edge lengths and extrinsic curvature dihedral angles \cite{tg,IoCarlo}. 
This relation between twisted geometry and Regge calculus implies that holonomies and fluxes carry \emph{more} information than the space of Regge calculus. This is not in contradiction with the fact that the Regge variables and the LQG variables on a fixed graph both provide a truncation of general relativity: simply, they define two distinct truncations of the full theory. 
See \cite{IoCarlo} for a discussion of these aspects.

For an arbitrary graph, the shape-matching subset describes a generalization of 3d Regge geometry to arbitrary cellular decompositions.
In this case however the variables are not equivalent any longer to edge lengths, since as already discussed these do not specify uniquely the geometry of polyhedra. Rather, such cellular Regge geometry must use areas and normals as fundamental variables. 
\\

Finally, let us make some comments on the coherent states themselves. The discussion so far is largely independent of the details of the coherent states on $\hh_\Gamma$. All that is required is that they are properly peaked on a point in phase space. 
The states most commonly used are the heat-kernel ones of Thiemann and collaborators.
Notice that these are not written in terms of the LS coherent intertwiners \Ref{CI}. Nevertheless, it was shown in \cite{BMP} that they do reproduce coherent intertwiners in the large area limit.
Alternative coherent states based directly on coherent intertwiners appear in \cite{Qtwi}.
These results show that coherent intertwiners can be used as building blocks of coherent spin networks.

\section{On the volume operator}\label{SecVol}

At the classical level, the volume of a polyhedron is a well-defined quantity. In this section we investigate the quantization of this quantity and its relation with the volume operators used in loop quantum gravity.

\subsection{The volume of a quantum polyhedron}

Let us consider the phase space $\mathcal{S}_F$ of polyhedra with $F$ faces of given area. The volume of the polyhedron is a well-defined function on this phase spase, as discussed in Section \ref{SecV}. Coherent intertwiners provide a natural tool to promote this quantity to an operator in $\hh_F$.

In the following we use the parametrization of the phase space $\mathcal{S}_F$ in terms of the cross ratios $Z_k$. In particular, the $F$ normals $n_i$ are understood as functions of the cross-ratios, $n_i(Z_k)$. Accordingly we call $V(j_i,Z_k)$ the volume of a polyhedron with faces of area $A_i(j_i)=8\pi\gamma L_P^2 j_i$ and normals $n_i(Z_k)$,
\begin{equation}
V(j_i,Z_k)\equiv V(A(j_i),n(Z_k))\;.
\label{eq:VZ}
\end{equation}  
For simplicity we assume an ordering of operators such that the area is linear in the spin, but the above expression, and the following
construction, can be immediately applied to other possibilities.

Let us consider now the Hilbert space of intertwiners $\mathcal{H}_F$ associated to the phase space $\mathcal{S}_F$. The volume of a quantum polyhedron can be defined in terms of coherent intertwiners $|j_i,Z_k\rangle$ and of the classical volume as follows:
\begin{equation}
\hat{V}=\int d\mu(Z_k)\;V(j_i,Z_k)\;|j_i,Z_k\rangle\langle j_i,Z_k|\;.
\label{hatV}
\end{equation}
This integral representation of the operator in terms of its classical version\footnote{In the literature \cite{Perelomov}, the classical function $V(j_i,Z_k)$ is called the $P$-symbol of the operator $\hat{V}$. On the other hand, the expectation value of the operator $\hat{V}$ on a set of coherent states, i.e.
\begin{equation}
Q(j_i,Z_k)\equiv \la j_i,Z_k|\hat{V}|j_i,Z_k\ra\;,
\nonumber
\end{equation}
is called the $Q$-symbol. When the $P$-symbol and the $Q$-symbol of an operator exist, then the operator is fully determined by either of them. The properties of these symbols and of the operator they define have been studied by Berezin in \cite{Berezin:1971jf,Berezin:1980xw}} is of the kind considered originally by Glauber \cite{Glauber:1963tx} and
Sudarshan \cite{Sudarshan:1963ts}. It has a number of interesting properties that we now discuss:
\begin{itemize}

	\item[Q1.] The operator $\hat{V}$ is positive semi-definite, i.e.
	\begin{equation}
\la\psi |\hat{V}|\psi\ra=\int d\mu(Z_k) \;V(j_i,Z_k)\;|\la j_i,Z_k|\psi\ra|^2\;\;\geq 0\;,
\label{eq:Vpos}
\end{equation} 
for every $|\psi\ra$ in $\hh$. This is a straightforward consequence of the fact that the classical volume is a positive function, $V(j_i,Z_k)\geq 0$. Furthermore, $\hat V$ vanishes for $F=2$ and $3$.

	\item[Q2.] $\hat{V}$ is a bounded operator in $\mathcal{H}_F$. Its norm $||\hat{V}||=\sup_{\psi}  \la \psi|\hat{V}|\psi\ra/\la \psi|\psi\ra$ is bounded from above by the maximum value of the classical volume of a polyhedron with fixed areas,
\begin{equation}
\frac{\la\psi |\hat{V}|\psi\ra}{\la\psi |\psi\ra}=\int d\mu(Z_k) \;V(j_i,Z_k)\;|\la j_i,Z_k|\psi\ra|^2\;\;\leq\;\; 
\sup_{Z_k} \left\{ V(j_i,Z_k) \right\} \;\equiv V_{max}(j_i)\;.
\label{eq:Vmaxq}
\end{equation}
\item[Q3.] \emph{0-spin consistency}. Let us consider the operator $\hat{V}$ defined on the Hilbert space $\mathcal{H}_{F+1}$ associated to spins $j_1,\ldots,j_F,j_{F+1}$, and the one defined on the Hilbert space $\mathcal{H}_{F}$ associated to spins $j_1,\ldots,j_F$. When the spin $j_{F+1}$ vanishes, the two operators coincide. This is a consequence of the fact that the classical volume of a polyhedron with $F+1$ faces coincides with the volume of a polyhedron with $F$ faces and the same normals when one of the areas is sent to zero. 
\end{itemize}
These three properties are the quantum version of C1, C2, C3 discussed in Section \ref{SecV}. Moreover, using the fact that for large spins 
two coherent intertwiners become orthogonal,
\begin{equation}
|\la j_i,Z_k|j_i,Z'_k\ra|^2\to \delta(Z_k,Z'_k),
\label{eq:deltaZZ}
\end{equation}
we have that the expectation value $\la \hat{V} \ra$ of the volume operator on a coherent state $|j_i,Z_k\ra$ reproduces the volume of the classical polyhedron with shape $(j_i,Z_k)$,
\begin{equation}
\la \hat{V} \ra\equiv\frac{\la j_i,Z_k|\hat{V}|j_i,Z_k\ra}{\la j_i,Z_k|j_i,Z_k\ra}\approx V(A_i(j_i),n_i(Z_k)).
\label{eq:Vcl}
\end{equation}
This fact allows to estimate the largest eigenvalue of the volume: in the large spin limit, the largest eigenvalue is given by $V_{max}(A_i)$, the volume of the largest polyhedron in $\mathcal{S}_F$.

\begin{figure}[h!]
\begin{center}
\includegraphics[scale=1.2]{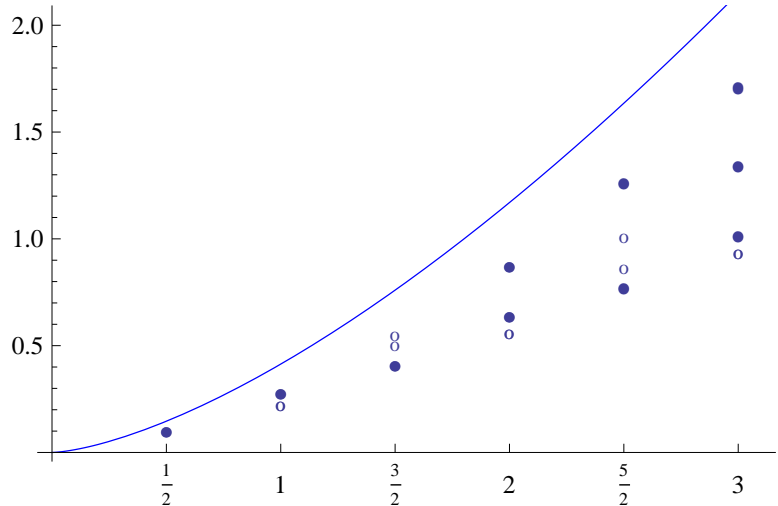} 
\caption{\small{\label{FigSpettro} Some eigenvalues of $\hat V$. For comparison, the curve is the classical volume of an  equilateral tetrahedron as a function of the area $A=j$ (units $8\pi \gamma L_P^2=1$). The empty circles are single eigenvalues, the full circles have double degeneracy. The spectrum is gapped and bounded from the above by the classical maximal volume, which provides a large spin asymptote.}}
\end{center}
\end{figure}

The spectrum of the operator $\hat{V}$ can be computed numerically. Let us focus on the case $F=4$ for concreteness. The matrix elements of $\hat{V}$ in the conventional recoupling basis are given by
\be
V_{kk'}=\bra{j_i,k}\hat{V}\ket{j_i, k'} = \int \rd \mu(Z) \,V(j_i,Z)\,  \la j_i,k \ket{j_i,Z} \bra{j_i, Z} j_i, k' \ra.
\ee
The matrix $V_{kk'}$ can be diagonalized numerically to obtain its eigenvalues.\footnote{
The overlaps, $\la j, k \ket{j_i,Z} = \frac{(-1)^{2k}}{2j + k +1} \frac{(2j!)^2}{(2j+k)! (2j-k)!} L_k(1-2 Z)$, where $L_k$ is the $k$-th Legendre polynomial, and the measure, can be found in \cite{CF3}.
} 
We focused for simplicity on the simplest case where all the four spins $j_i$ are equal to $j_0$. The results using Wolfram's Mathematica are shown in Fig. \ref{FigSpettro} and confirm that the maximum eigenvalue is below the volume of the regular tetrahedron.  
Notice also that the spectrum has a gap. One of the interesting questions to investigate in the future is whether this gap survives at higher valence, or it decays as for the standard volume operator \cite{Brunnemann}.

It is interesting to notice that the volume operator introduced above commutes with the parity operator. This is the operator that sends the normals to their opposite,
\be 
\hat{ \mathcal P} \ket{j,n} = \ket{j,-n}.
\ee
In terms of the stereographic projection, the maps $n\mapsto -n$ amounts to $z\mapsto -1/\bar z$, thus its action on coherent intertwiners labeled by the single cross-ratios $Z$ is simply
\be 
\hat{ \mathcal P} \ket{j_i,Z} = \ket{j_i,\bar Z}.
\ee
Notice that $V(j_i,Z_k)=V(j_i,\bar Z_k)$ thanks to the invariance of the classical volume under parity. Moreover the measure $\rd\mu(Z_k)$ is invariant under the transformation $Z_k\to \bar{Z_k}$. As a result, the operator \Ref{hatV} commutes with parity,
\begin{equation}
\hat{ \mathcal P} \hat V \hat{ \mathcal P}^\dagger = \int \rd\mu(Z)\, V(j_i,Z) \ket{j_i,\bar Z}\bra{j_i,\bar Z} 
=\int \rd\mu(\bar Z)\, V(j_i,\bar Z)\ket{j_i,Z}\bra{j_i,Z} = \hat V.
\end{equation}
This explains the degeneracies seen in the spectrum.

\bigskip

Clearly, there are other possibilities for the volume of a quantum polyhedron. All of them share the same classical limit, but can have a different spectrum for small eigenvalues. An interesting variant is $\hat{\widetilde V}=\sqrt{|\hat{U}|}$, where $\hat U$ is the oriented-volume square operator, defined as
\begin{equation}
\hat{U}=\int d\mu(Z_k)\;s(Z_k) V^2(j_i,Z_k)\;|j_i,Z_k\rangle\langle j_i,Z_k|.
\label{eq:Uhat}
\end{equation}
Here $s(Z_k)$ is the parity of the polyhedron, i.e. $s(Z_k)=\pm 1$ and $s(\bar{Z}_k)=-s(Z_k)$. 

\begin{figure}[h!]
\begin{center}
\includegraphics[scale=.8]{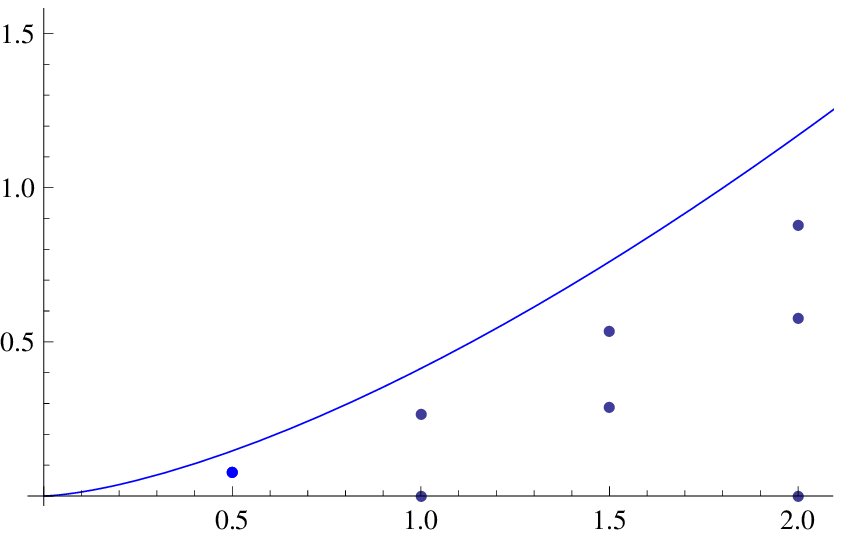} \hspace{1cm}
\includegraphics[scale=.8]{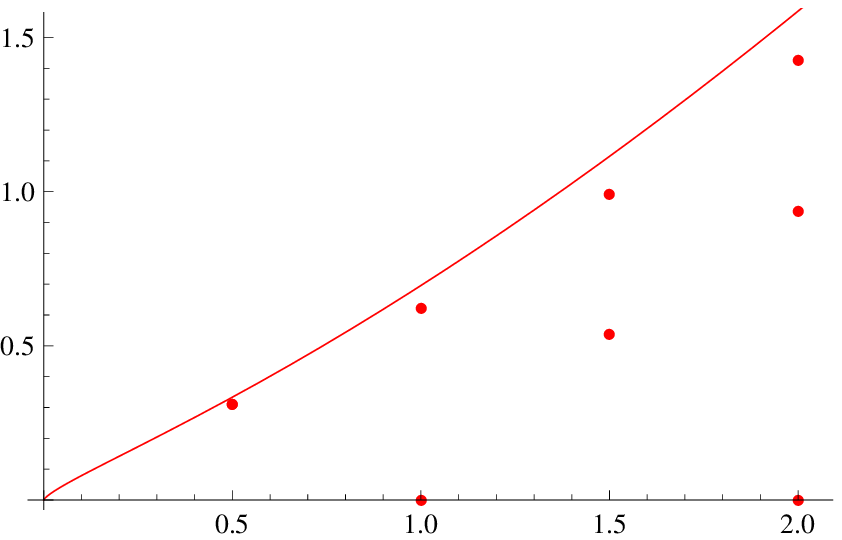} \caption{\small{\label{FigSpettroOr} Left panel. Some eigenvalues of $\hat{\widetilde V}$. For comparison, the curve is the classical volume of an  equilateral tetrahedron as a function of the area $A=j$ (units $8\pi \gamma L_P^2=1$).
All but the zero eigenvalue have double degeneracies.
Right panel. Same region of the spectrum for Barbieri's operator $\hat{V}_B$. Notice that here the asymptotic curve is the equilateral volume with areas $A=\sqrt{j(j+1)}$.
}}
\end{center}
\end{figure}

The operator $\hat{U}$ anticommutes with the parity, and so does $\hat{\widetilde V}$. Therefore, under the assumption that the spectrum is non-degenerate, we have that the eigenvalues appear in pairs $\pm u$. In particular, a zero eigenvalue is present when the Hilbert space $\mathcal{H}_F$ is odd-dimensional.
This operator is similar in spirit to the the volume of a quantum tetrahedron introduced by Barbieri \cite{Barbieri}, 
 $\hat{V}_B=(8\pi \gamma)^{\frac{3}{2}} L_P^3 \frac{\sqrt{2}}{3}\sqrt{|J_1\cdot(J_2\times J_3)|}$. In Fig. \ref{FigSpettroOr} we show some eigenvalues of $\hat{\widetilde V}$ and a comparison with $\hat{V}_B$. 
 
For more on semiclassical aspects of the spectrum of the volume, see \cite{Hal}.
 
\subsection{LQG volume operator and the quantum polyhedron}

In LQG, the operator associated to the volume of a region in space is a well studied quantity \cite{RS,AL,Volume}. It is defined on the graph Hilbert space $\hh_\Gamma$ as a sum over contributions $\hat{V}_n$ from each node $n$ of the graph within the region $R$,
\begin{equation}
\hat{V}_\Gamma(R)=\sum_{n\subset R}\hat{V}_n.
\label{eq:VGamma}
\end{equation} 
In order to admit a lifting from $\hh_\Gamma$ to the full Hilbert space of LQG, the operator $\hat{V}_\Gamma(R)$ has to satisfy a number of consistency conditions that go under the name of ``cylindrical consistency'' \cite{ThiemannBook}. In particular, these conditions are satisfied by the operator $\hat{V}_n$ if $(i)$ it commutes with the area of dual surfaces, so that $\hat{V}_n$ reduces to an operator on the intertwiner space $\hh_{F(n)}$, and $(ii)$ it satisfies a \emph{0-spin consistency condition} 
so that the operators defined on different intertwiner spaces coincide when these spaces are identified.

\medskip

In the previous section we have introduced an operator $\hat{V}_n$, given by \Ref{hatV} for the given node, that satisfies these conditions. Condition $(i)$ holds because by construction the operator acts within $\hh_{F(n)}$, and condition $(ii)$ follows from property Q3 in Section \ref{SecVol}.
This operator is based on the knowledge of the classical system behind the intertwiner space $\hh_{F(n)}$. The single node operator $\hat{V}_n$ measures the volume of a quantum polyhedron dual to the node, and the operator $\hat{V}_\Gamma(R)$ built as in \Ref{eq:VGamma} the volume of a region in a twisted geometry. It has a good semiclassical limit by construction.

\medskip

The standard strategy in LQG is on the other hand rather different. The starting point is the classical expression for the volume of a region,
\begin{equation}
V(R)=\int_R\rd^3 x \,\sqrt{\frac{1}{3!}\big|\epsilon^{ijk}\epsilon_{abc}E^a_i E^b_j E^c_k\big|}\;,
\label{eq:Vgr}
\end{equation}
$E^a_i(x)$ being the Ashtekar-Barbero triad. The key step is to rewrite this quantity in terms of fluxes, which are the fundamental operators of the theory. This step introduces a regularization procedure which is
adapted to a graph $\Gamma$ embedded in space. Then, the regularized quantity is promoted to an operator in the Hilbert space $\hh_\Gamma$ and the limit of vanishing regulator exists and it is well-defined. Two volume operators have been constructed in this way, one by Rovelli-Smolin \cite{RS}, and one by Ashtekar-Lewandowski \cite{AL}. Both these operators have the form (\ref{eq:VGamma}), and differ in the regularization procedure and in details on the exact form of $\hat{V}_n$. For the Ashtekar-Lewandowski volume operator, the node contribution is defined on the intertwiner space $\hh_F$ as
\begin{equation}
\hat{V}_n^{AL}=(8\pi \gamma)^{3/2} L_P^3\, \sqrt{\frac{1}{8}\Big|\sum_{1\leq i<j<k\leq F}\!\!\!\epsilon(e_i,e_j,e_k) \;\vec{J}_i\cdot(\vec{J}_j\w\vec{J}_k)\Big|},
\label{eq:VAL}
\end{equation}
where $\epsilon(e_i,e_j,e_k)=\pm 1, 0$ is the orientation of the tangents $e_i$ to the links at the node. The overall coefficient is fixed by a consistency requirement known as `triad test' \cite{Giesel}. There is a large amount of analytical and numerical results on the spectrum of this operator (e.g. \cite{Volume,Brunnemann}), particularly because it enters Thiemann's construction of the Hamiltonian constraint \cite{ThiemannHam} and thus it is relevant to understand the quantum dynamics of the theory. Moreover its semiclassical behaviour has been investigated in detail with the conclusion that only \emph{cubulations}, that is regular graphs with $6$-valent nodes, have a good semiclassical limit \cite{Flori}. In the light of the quantum polyhedron introduced in this paper, this result can be understood as follows.

On semiclassical states,\footnote{The semiclassical states used in the analysis of \cite{Flori} are the heat-kernel coherent states developed by Thiemann and collaborators \cite{Thiemann}. However, the details on the coherent states do not matter for our argument, all that is required is that they are peaked on a given point in the classical phase space $S_\Gamma$.} 
 $\mean{\vec J_i} = \vec{A}_i\equiv A_i n_i$ (see discussion in Section \ref{SecLQG} and cf. \Ref{eq:corresp} and \Ref{eq:jn}), and the expectation value of \Ref{eq:VAL} is -- at zero order in $\hbar$ \cite{Flori}
\begin{equation}
\mean{\hat{V}_n^{AL}}=\sqrt{\frac{1}{8}\Big|\sum_{1\leq i<j<k\leq F}\!\!\!\epsilon(e_i,e_j,e_k) \;
\vec A_i\cdot(\vec{A}_j\w\vec{A}_k)\Big|}.
\label{eq:VALev}
\end{equation}
As discussed earlier, the variables $\vec{A}_i$ of the semiclassical state define a polyhedron around the node $n$. The key observation is that \Ref{eq:VALev} is \emph{not} the volume of that polyhedron.
The volume of a convex polyhedron with $F$ faces is in general a rather complicated function of the areas and normals (see the discussion in Section \ref{SecV}). There is however a case where this expression simplifies greatly, and in this case it coincides with \Ref{eq:VALev}: it happens for \emph{parallelepipeds}. Parallelepipeds are a subset of the phase space $\cs_F$ for $F=6$ with areas that are equal in pairs. They live within the combinatorial class of cuboids: they are cuboids with three couples of parallel faces.\footnote{Notice that parallelepipeds are a set of measure zero among the cuboids. Moreover, cuboids are not the only dominant class in phase space $\cs_F$ with $F=6$.} The volume of a parallelepiped is 
\be
{V}=\sqrt{ | \vec A_1\cdot(\vec{A}_2\w\vec{A}_3) |},
\label{Vcubo}
\ee
where $(123)$ are any three faces sharing a vertex. It is straightforward to see that this coincides with \Ref{eq:VALev} for the semiclassical state of a cubic analytic node\footnote{That is, the link are the analytic continuations of each other across the nodes.} with areas equals in pairs and normals parallel pairwise.

This fact explains why the expectation value of the operator \Ref{eq:VAL} on a semiclassical states reproduces the volume of a parallelepiped for $F=6$, but not the volume of other polyhedra.\footnote{It goes without saying that the dependence on areas and normals of the expression \Ref{eq:VAL} can be used to define the volume of a tetrahedron, as we saw with $\hat V_B$ earlier. But that would require a different numerical coefficient in \Ref{eq:VAL} -- an extra $\sqrt{2}/3$ -- which is hard to motivate in the standard LQG construction.}

\section{On dynamics and spin foams}\label{SecSF}

Spin foam models for the dynamics of loop quantum gravity are usually built starting from a discretization of the spacetime manifold in terms of a simplicial triangulation $\Delta$. 
A certain control over the dynamics comes from a connection with Regge calculus in the large spin limit. Specifically, in this limit the  transition amplitudes are related to exponentials of the Regge action \cite{CF3,Barrett,CF,BarrettLor}.
This result is generally regarded as a promising step towards understanding the low-energy physics of the theory, since discrete general relativity on $\Delta$ is reproduced.
On the other hand, complete transition amplitudes for LQG require the use of more general 2-complexes than those those dual to simplicial manifolds.\footnote{Although a direct construction of the path integral for arbitrary graph has not been attempted so far, in \cite{KKL} a model valid for arbitrary graph was proposed, based on a natural extension of some algebraic properties of the EPR model \cite{EPR}.}

Just as Regge calculus is useful to study the semiclassical behaviour on simplicial manifolds, a generalization thereof
to arbitrary cellular decompositions could be relevant to the full theory, and allow us to test whether models such as the one proposed in \cite{KKL} can be related to (discrete) general relativity. 
In this final Section, we would like to make two remarks on this idea.

The first remark concerns Regge calculus on arbitrary cellular decompositions. 
The point is that edge lengths are not good variables to capture the (discrete) metric of the manifold. 
This is simply because a generic 4d polyhedron at fixed edge lengths is not rigid. Therefore a piecewise-linear metric can not be described by the edge lengths of the polyhedra alone.
The solution to this problem can be found looking again at Minkowski's theorem, which holds in any dimension. The theorem implies that a generic polyhedron in $\R^n$, sometimes called an $n$-polytope, is uniquely characterized by $nF-n(n+1)/2$ numbers: the volumes of the $F$ ``faces'' (which are now $(n-1)$-polytopes) and the normals satisfying the $n$-dimensional closure condition. 
On the other hand, $n$-simplexes are polytopes with a minimal number of faces, $F=n+1$. In this case, assigning their $n(n+1)/2$ edge lengths suffices, thus edge lengths fix a unique flat metric on each $n$-simplex and can be used as fundamental variables in the full triangulation. 

Let us fix $n=4$. To identify the geometry of each 4-polytope, we need volumes $V_m$ and 4d unit normals $N_m$ of each polyhedron $m$ in its boundary, satisfying the closure condition. For these to extend to a piecewise-linear, continuous metric on the whole cellular decomposition, we additionally need shape matching conditions, of the sort described in Section \ref{SecSM} for three dimensions. 
A tentative Regge-like action can then be written as
\be\label{ReggeOnSt}
S[V_m, N_m] = \sum_f A_f(V_m,N_m) \eps_f(V_m, N_m) + {\rm constraints},
\ee
where $f$ are the 2d faces of the cellular decomposition, and $\eps$ the deficit angles, defined as usual as $2\pi$ minus the sum of dihedral angles of each 4-polytope sharing the face. The constraints are the closure and shape matching conditions.
In principle, we can interpret \Ref{ReggeOnSt} as an ``effective'' Regge action in which the internal edge lengths of an initial simplicial triangulation have been evaluated on the flat solution.

The second remark concerns the link between spin foam amplitudes and Regge calculus.
A lesson from the recent asymptotics studies of the EPR model is that the amplitude is dominated by exponentials of the Regge action when the boundary data satisfy certain conditions, which guarantee the existence of a unique 4-simplex in the bulk. This suggest that the dominant contributions to models on arbitrary graphs could come from requesting the existence of a unique 4-polytope, and that the amplitude could be related to a form of the Regge action specialized to the 4-polytope, such as the one described above. 
So the question is whether, as for the 4-simplex, the conditions for the existence of the 4-polytope can be mapped into conditions on the boundary data, such as 3d closure and non-degeneracy conditions, and shape matching. This is a key question that we leave open for future work. We believe that the answer, and these considerations in general, will be relevant to tackle the problem of the semiclassical limit of spin foams on arbitrary graphs, such as the one proposed in \cite{KKL}.

\section{Conclusions}\label{Conclusions}

In this paper we discussed a number of properties of classical polyhedra which are of interest to loop quantum gravity.
A polyhedron can be uniquely identified by the areas and the normals to its faces (Minkowski's theorem \cite{Minkowski}, Section \ref{phase space}). The identification includes the knowledge of its geometry (edge lengths, volume), and its combinatorial class (the adjacency of the faces). This information can be explicitly derived from the areas and normals through the reconstruction procedure presented in section \ref{SecRec}. We observed that the space of polyhedra of given areas is a phase space, previously introduced by Kapovich and Millson \cite{Kapovich}, and used our reconstruction algorithm to divide this space into regions corresponding to different classes. 

We then discussed the relevance of polyhedra to the quantum theory. We first recalled that the quantization of Kapovich and Millson phase space gives the SU(2)-invariant space of intertwiners (section \ref{SecLQG}), and thus observed that the LS coherent intertwiners can be interpreted as semiclassical polyhedra. The polyhedral picture can be extended to a whole graph using the twisted-geometry parametrization of the holonomy-flux variables introduced in \cite{tg}.
The knowledge of the classical space behind intertwiners was then used to introduce a new operator, which measures the volume of a quantum polyhedron (section \ref{SecVol}), and by construction has the correct semiclassical limit. We performed some numerical analysis of its spectrum for the simplest 4-valent case.
We discussed its relation to the volume operators commonly used in loop quantum gravity.
Finally (section \ref{SecSF}), we used the four-dimensional version of Minkowski's theorem to make some remarks on Regge calculus on non-simplicial discretizations and its possible relevance to spin foam models on graphs of arbitrary valence.

Our hope is that the notion of a quantum polyhedron can find useful applications in future developments of loop quantum gravity, and that the results in this paper are a first step in that direction.

\section*{Acknowledgments}
The authors are grateful to Hal Haggard and Carlo Rovelli for many useful discussions and for comments on a first version of this paper. The work of E.B. is supported by a Marie Curie Intra-European Fellowship within the 7th European Community Framework Programme. The work of S.S. is partially supported by the ANR ``Programme Blanc'' grant LQG-09.



\end{document}